\documentclass{aa}

\input epsf.sty

\usepackage{lscape}

\begin{document}

\def\Agata{R\' o\. za\' nska~}
 \def\Ka{K$_{\alpha}$}

               
%
   \title{Modeling the Warm Absorber in AGN}

   \author{A. \Agata\inst{1}
        R. Goosmann\inst{2}
        A.-M. Dumont\inst{2} 
	 B. Czerny \inst{1}
          }

   \offprints{A. \Agata (agata@camk.edu.pl)}

   \institute{
   $^1$Copernicus Astronomical Center, Bartycka 18, 00-716 Warsaw, Poland \\
     $^2$Observatoire de Paris-Meudon, LUTH, Meudon, France \\
             }
  \authorrunning{}
  \titlerunning{}
   \date{Received ...; accepted ...}

   \abstract{We present a wide grid of models for the structure and transmission
properties of warm absorbers in active galactic nuclei (AGN). 
Contrary to commonly used constant density models,
our absorbing cloud is assumed to be under {\it constant total (gas plus radiation) 
pressure}. This assumption implies the coexistence of  material at different 
temperatures and ionization states, which is a natural consequence of
pressure and thermal equilibrium. 
Our photoionization code allows us to 
compute the profiles of the density, the temperature, the gas pressure, 
the radiation pressure
and the ionization state across the cloud,
and to calculate  the radiative transfer of continuum and lines 
including Compton scattering.
Therefore, equivalent widths 
of both saturated and unsaturated lines are properly modeled.
For each pair of the incident spectrum slope and the ionization parameter
at the cloud surface there is a natural upper limit to the total column densities
of the cloud due to thermal instabilities. 
These maximum values are comparable to the observational
constraints on the column density of warm absorbers which may
give support to constant total pressure models.
In all models we note considerable absorption around 6.4 keV
which modifies the intrinsic 
relativistically broadened iron line profile originating
in an accretion disk illuminated atmosphere.
Our models can be applied to fitting the spectroscopic data from 
the {\it XMM-Newton} and {\it Chandra} satellites.

}
\maketitle
%

\section{Introduction}

The first X-ray absorption feature due to ionized heavy elements 
was recognized by Halpern (1984) in the X-ray spectrum of the Sy1 galaxy
MR 2251-178 observed by the {\it EINSTEIN} satellite. 
The author related the observed jump  in flux around 1 keV to the  
absorption edge of O{\sc viii}. Therefore, as
Halpern concluded, X-rays emitted from the central region of an active galaxy
traveling toward an observer encounter a 
``warm absorber'' --- material with  an
electron temperature lower than the temperature 
of collisionally ionized gas with a similar level 
of ionization.   

Results from {\it EXOSAT}, {\it ROSAT}, {\it GINGA}, 
{\it ASCA}, and {\it Beppo-SAX} satellites showed that 
warm absorbers were common among Seyfert galaxies. 
Nandra \& Pounds (1994) have suggested
that more than 50\% of Sy1s contain a warm absorber, and
later results confirmed such occurrence rate (e.g.
Reynolds, 1997; George at al. 1998).
Those conclusions were based exclusively on the detection of
absorption edges (for instance in MCG-6-30-15, Nandra \& Pounds 1992).
The detector areas and the spectral resolution of those satellites did not 
allow us to see any absorption lines from  highly ionized species. 

The situation changed since 1999, when large X-ray telescopes 
{\it Chandra} and {\it XMM-Newton} started to operate, with 
their X-ray grating 
instruments working in the energy range up to almost 10 keV.
Several tens of absorption or emission lines were 
observed and identified in NGC 3783 (Kaspi el al. 2002, Behar el al. 2003,
Netzer et al. 2003, Krongold el al. 2003), NGC 5548 (Kaastra el al. 2002), 
NGC 1068 (Kinkhabwala et al. 2002), NGC 7469 (Blustin et al. 2003),
 MCG-6-30-15 (Turner et al. 2004).
In other objects the number of fitted lines is lower and/or their detection
is less firm, but the  results 
still strongly support the presence of a warm absorber in those Seyferts
(NGC 4051, Collinge et al. 2001; Mrk 509, Yaqoob el al. 2003;
TonS180, \Agata el al. 2004). Warm absorber lines were even detected in
a distant blazar at $z = 4.4$ (Worsley et al. 2004).

Detected lines are basically consistent with the unification scheme
of AGN based on the presence of the dusty/molecular torus. 
According to that scheme, Sy1 are objects seen face-on and 
Sy2 are objects seen edge-on
(Antonucci \& Miller 1985; for application to X-ray band see Mushotzky,
Done \& Pounds 1993). 
The major characteristic of the X-ray spectra of Sy2 galaxies
is that they are strongly absorbed at low energies, and that their
emission lines, especially the iron K$\alpha$ line, have large equivalent
widths (Turner et al. 1997) since their intensity 
is measured with respect to the heavily obscured
direct continuum (Weaver \& Reynolds 1998) or only with respect to the 
scattered 
continuum (in Compton-thick objects), as discussed by Bassani et al. (1999).
In some cases of Compton-thin Sy2 obscuration may be due to the
host galaxy and unrelated to disk/torus orientation (Matt 2000, 
Guainazzi et al. 2001)
The spectra of Sy1 galaxies are predominantly featureless,
show little low energy absorption,
and the equivalent width of the iron K$\alpha$ line is much lower
than in Sy2. Some Sy1 show relativistically broadened iron K$\alpha$ line
profiles which additionally supports the view that in Sy1 we directly observe
the innermost part of the nucleus. The scattering medium may be identical
to the warm absorber and we have side view of this medium in Sy2 galaxies
while we see the nucleus through it in Sy1. Indeed,
X-ray spectra of some Sy2 galaxies show many narrow emission lines,
(e.g. NGC 1068, Kinkhabwala et al. 2002; NGC 4507, Matt et al. 2004; 
Mkn 3, Pounds \& Page 2005; NGC 4151, Schurch et al. 2004) indicating that 
ionized material may extend beyond the shielding torus.
In Sy1 spectra mostly absorption 
lines are observed (e.g. NGC 3783, Kaspi et al. 2002: NGC 5548, 
Kaastra et al. 2002; Mrk 509, Yaqoob et al. 2003) without
strong absorption of continuum, 
suggesting that in those objects we see directly the nucleus through the 
ionized plasma.  Some
contribution of emission lines in Sy1 X-ray spectra
is also expected (e.g. Netzer 1993, Collin, Dumont \&
Godet 2004) and seen in the data (e.g. NGC 3783, Behar et al. 2003; 
NGC 7469, Scott et al. 2005). However, the constraints
from Sy2 galaxies for the ionized medium cannot be used directly to Sy1 
galaxies since in Sy2 galaxies we may observe only the outer part of the
plasma distribution, due to the obscuration by the torus, 
while warm absorber features
may come predominantly from the inner part.

A majority of these absorption lines shows a velocity shift of the order 
of a few hundreds
km/s (Kaspi et al. 2001, Kaastra et al. 2002) suggesting that the 
warm absorber is outflowing.
A strong high velocity outflow was recently reported in several objects,
mostly radio-quiet quasars 
(e.g. 60000 km/s and 120000 km/s in APM 08279+5255, Chartas et al. 2002; 
23000 km/s in PG1211+143, Pounds et al. 2003a;
63000 km/s in PG0844+349, Pounds et al. 2003b, 30000 km/s in IRAS 13197-1627,
Dadina \& Cappi 2004; 26,000 km/s in PG 1404+226, Dasgupta et al. 2005).
However, the results are based on a few detected lines,
so line identifications and, consequently, the determined high outflow 
velocities
can be questioned (Kaspi 2004a). 
The origin of the outflow and its geometry is still under discussion
(Crenshaw, Kraemer \& George 2003a, Blustin et al. 2005). 

It is more difficult to estimate the radial distance of the warm absorber 
from the nucleus.
Absorption lines most probably form somewhere between 
the broad line region (BLR) and   the narrow line region (NLR)
(i.e. from about 0.01-0.1 up to 10 pc from nucleus; 
see Crenshaw et al. 2003a, Blustin et al. 2005).
Variability in the overall warm absorber properties
was reported for a few sources (MR 2251-178, Halpern 1984, 
Kaspi et al. 2004b; MCG -6-30-15, 
Reynolds et al. 1995;
H1419+480, Barcons et al. 2003; NGC 4395, Shih et al. 2003; NGC 3516,
Netzer et al. 2002).
Netzer et al. (2002) and Barcons et al. (2003) concluded that the changes observed
are consistent with varying ionization of the gas.  
The lack of short-timescale (days) response of the warm absorber to
the change of the continuum can be used to put lower limits to the
warm absorber distance (e.g. 0.5 - 2.8 pc for NGC 3783, Behar et al.
2003; similar limits were given by Netzer et al. 2003).
Krongold et al. (2005) detected a response of the warm absorber to 
the changes of the continuum on a timescale of 31 days in NGC 3783,
deriving an upper limit of 6 pc for the distance of the warm absorber.
The shortest variability timescale of $\sim 10^4$ s has been detected in
the warm absorber in MCG -6-30-15 (Otani et al. 1996; see also Turner et al.
 2004) locating the plasma responsible for the O{\sc viii} edge within
the distance of $10^{17}$ cm from the nucleus.  
The spectral analysis of this source indicates that the highly ionized 
warm absorber is dust-free, with dust contribution in this source coming from 
a distant zone, hundreds of pc from the nucleus (Ballantyne et al. 2003).  

The same medium is most probably
responsible for narrow absorption lines seen in the UV spectra of many
AGN (for a review, see Crenshaw et al. 2003a). Some kinematic components 
discovered in UV coincide with those discovered in soft X-rays 
but
for other components no such correspondence is seen (e.g. Behar et al. 2003 
for NGC 3783, Crenshaw et al. 2003b for NGC 5548, Kaspi et al. 2004b for 
MR 2251-178; Scott et al. 2005 for NGC 7469; Gabel et al. 2005 for NGC 3783). 
However, the resolving power
of the UV observations is high ($R \sim 20 000$) while X-ray data
resolution is much lower ($R \sim 1 000$), which makes the comparison
difficult, as discussed by Crenshaw et al. (2003a). Analysis usually
suggests that UV and X-ray absorption features are consistent with arising
in the same gas, but with stratified ionization (e.g. Barcons et al. 2003,
Kaspi et al. 2004b, Scott et al. 2005, Gabel et al. 2005). 
Since the absorption features appear in the profiles
of the broad emission lines like C{\sc iv}, and are occasionally deep, this
serves as an argument that the absorbing region is located outside the
BLR.

The column density of the warm absorber is generally estimated to be about 
$10^{21-23}$ cm$^{-2}$, and absorbing gas contains heavy elements
mostly in the form of helium- and hydrogen-like ions. However, accurate
measurements of the column density are quite complex. Most determinations
are based on detection of absorption edges, but in some cases 
edges are undetectable while lines are clearly seen 
(Kaastra el al. 2002, \Agata et al. 2004). Also the estimates 
of column densities from edges do not always 
confirm estimates derived from the absorption line analysis 
(Kaspi et al. 2002).

The ionization state of the gas required to reproduce the observed
absorption or emission lines seems to be quite complex.  
In many objects predictions based on a single cloud at one specific 
ionization state cannot explain the data, 
so an absorbing material is modeled 
using at least two photoionization regions, which  are 
required to explain presence of lines from the matter in 
different ionization states 
(see in the case of Sy1:  NGC 4051, Collinge et al. 2001; 
NGC 5548,  Kaastra et al. 2002;  NGC 3783,  Kaspi et al. 2002, 
Netzer et al. 2003 and
Krongold et al. 2005, H0557-385, Ashton et al. 2005). 
The same conclusion was drawn from fitting of the absorbed continuum
(strongly absorbed Sy1: Mkn 304, Piconcelli et al. 2004; IC 4329A, 
Steenbrugge et al. 2005a,b;
in the case of a dwarf galaxy with an active nucleus: NGC 4395, 
Shih, Iwasawa \& Fabian 2003; 
and in case of Sy2: NGC 4507, Matt et al. 2004).


The idea of the warm absorber being under the constant pressure 
was developed even before {\it Chandra} and {\it XMM-Newton}
satellites were lunched  (Netzer 1993; Krolik \& Kriss 1995; Netzer 1996). 
It is well known that cold/warm material irradiated by hard  X-rays
should be strongly stratified, and eventual thermal instabilities
lead to its clumping (McKee Krolik \& Tarter 1981).
If thermal instabilities are strong, none of 
the existing photoionization codes  can describe unstable 
zone. This is because radiative transfer codes are based
on unique density and temperature profiles, and do not
accept situations where for one value of  optical depth 
the solution gives three different values of temperatures and densities. 

Therefore many models of the warm absorber being under
constant pressure do assume that those two discrete phases already exist.
Usually two or three zones at different constant densities
are assumed to have the same dynamical ionization parameter, $\Xi$, which is the
ratio of ionization pressure to the gas pressure (McKee Krolik \& Tarter 1981).
In such situation, calculations of transfer
of X-ray radiation through two constant density zones are done separately and 
then spectrum is merged together depending on covering 
factor  (Netzer 1993).

In this paper we solve the situation where X-ray radiation is 
no so hard and strong, that separation on two phases takes place.
Instead, our warm absorber is strongly stratified and radiation passes
through different densities and ionization stages. Theoretically 
this situation was considered by  
Krolik \& Kriss (2001), Krolik (2002). The stratification of 
our warm absorber is determined physically by radiative properties, and 
the only assumption which we make is constant total pressure within a 
cloud.  
We use the photoionization code {\sc titan} developed by Dumont et al. (2000)
(see Dumont et al. 2003 for implementation of Accelerated Lambda Iteration
method)
to compute the synthetic spectra for a systematic set of model
parameters. 
The advantage of our calculations is that we compute 
the full radiative transfer of continuum and lines
taking care also of lines which are  saturated.
Recent observations suggest that saturated lines
are often present in warm absorber
(Kaspi et al. 2002, \Agata et al.2004). The transfer is done
in the stratified medium, and the density profile is determined
self-consistently with the radiation transfer to fulfill the
condition of the pressure equilibrium across the cloud.
We aim understanding these AGN which
show clear absorption lines in their {\it Chandra} or {\it XMM} spectra 
so the warm absorber is located in the line of sight to the observer. 
We concentrate on modeling the transmission spectra and absorption lines
therefore our models can be used for Sy1 galaxies. 

We describe our model in Sec.~\ref{sect:model} and ~\ref{sec:grid}, 
while results are presented
in Sec.~\ref{sect:struc}, ~\ref{sect:local_spot}, and \ref{sec:rel}.
Sec.~\ref{sec:comp} describes
comparison of our models
to observations and  Sec.~\ref{sec:disc} contains conclusion remarks.

\section{Description of the model}
\label{sect:model}

The main assumption of our model is that the structure of a single warm absorber 
cloud is determined by the  condition of constant total, $P_{gas}+P_{rad}$, 
pressure.
The value of this pressure is self-consistently determined
by solving non-LTE ionization equilibrium, thermal equilibrium  and radiative 
transfer throughout the cloud thus providing a density and 
ionization profile inside a cloud.

We model an irradiated cloud using a plane parallel approximation. 
The total column density of the cloud, $N_H$, is a free parameter in our model.

The second model parameter is the ionization parameter, $\xi$, determined
at the cloud surface
\begin{equation}  
\xi=\frac{L}{n_0 R^2}=\frac{4 \pi F}{n_0},   
\end{equation}
where $L$ is the luminosity of the central source, $n_0$ is the number density 
at the cloud surface, at its illuminated side, and $R$ is the 
distance of the cloud from the central source. 
There is no difference if we use small $\xi$ or  dynamical 
big $\Xi$ ionization parameter which is defined as $\Xi=F/(cP_{gas})$ 
since both parameters are used only to specify the
amount of X-ray flux illuminating the cloud surface. In our models
both ionization parameters change with the optical depth 
due to the change in the density and in the irradiating flux, and 
this stratification is computed self consistently. 

The X-ray illuminating continuum is assumed to have a power-law  shape,
as it is usually observed in many Seyferts, and the power-law photon index, 
$\Gamma$, is a model parameter.

The photoionization code {\sc titan} developed by Dumont et al. (2000)
solves the full radiative transfer of the ionizing continuum in two stream
approximation. 
The transfer method for the lines and continuum used in this code
has been modified to use the accelerated lambda iteration (ALI) method,
which is especially well suited to rapid convergence in the cores
of lines which may be optically thick (Dumont et al. 2003). 
The computations are done assuming complete redistribution 
function in the lines.
Partial redistribution is mimicked by a Doppler profile for
 some of the most intense
resonant lines. 

The comparison of proper ALI computations with escape probability 
approximation was shown in the recent papers by Dumont et al. (2003) and 
Collin, Dumont, Godet (2004). In the case of emission lines 
the errors in the line fluxes are typically of the order of 30\% for 
column densities $10^{20}$ cm$^{-2}$ and 
a factor of five for column densities of $10^{23}$ cm$^{-2}$, 
compared to calculations using the escape probability formalism.

{\sc titan} includes all relevant physical processes from each ion level.
The population of each level is computed solving the set of 
ionization equations coupled with the set of statistical 
equations describing the excitation equilibrium. However, some low ionization 
states are more roughly treated.
 
The following processes
are taken into account: radiative and collisional ionization in all levels,
recombinations in all levels, and radiative and collisional excitations 
and de-excitations for all transitions. 

We consider ten main heavy elements including iron. 
{\sc titan} transfers ``only'' 900 lines 
while some photoionization codes ({\sc xstar}, {\sc cloudy})
collect over 4000 lines or more. 
On the other hand, {\sc titan} is also designed to
calculate radiative transfer in Compton thick media, which basically means that we
are able to compute transfer of optically thick lines
through any stratified cloud, and also through an
atmosphere in hydrostatic equilibrium (\Agata et al. 2002).
The same code is applicable for a whole range of
column densities (from small to large),
and a whole range of ionization states. 

The Compton heating/cooling 
balance through the cloud is taken into account since
for hard photon index, $\Gamma=1.5$, Compton heating is 
comparable to photoionization heating. 
Compton balance of photons up to 26 keV is computed in {\sc titan} 
while Compton balance of photons between 26 up to 100 keV 
is computed using code {\sc noar} (Dumont, Abrassart, \& Collin  2000).
The {\sc noar} is a radiative transfer code, based on the Monte Carlo
method, to compute a Compton heating profile for a plane stratified
atmosphere. This Compton heating/cooling profile is
read by the {\sc titan} and taken into consideration for solving the
radiative transfer. Usually three iterations between {\sc noar} and {\sc titan}
are enough to determine Compton heating/cooling curve. 
For the sets of our models with the same $\Gamma$ the Compton
heating does not vary much with $\xi$ and $N_{H_{tot}}$ 
of the warm absorber. Hence, we only derived a few representative
Compton profiles, for different $\Gamma$ and we adopted them to 
all other models to perform our computations. 
Comptonization of line photons is included.

\begin{figure}
\epsfxsize=8.8cm \epsfbox[20 150 600 700]{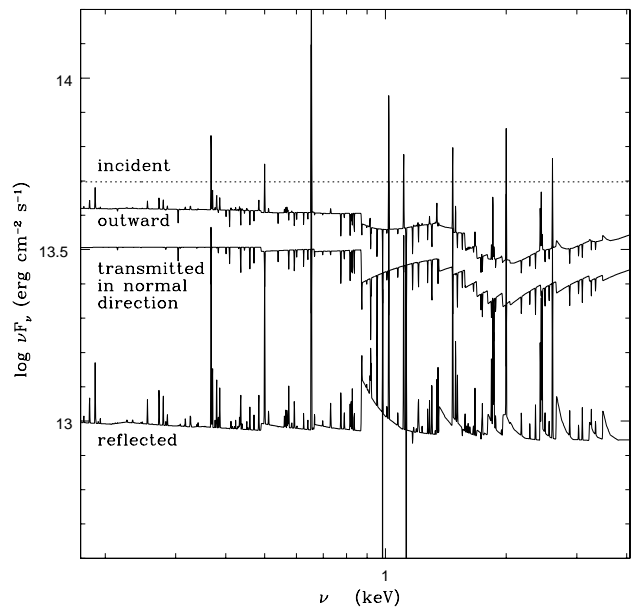}
\caption{All spectra available from one  considered cloud 
($\xi=10^5$, $N_{tot}=10^{24}$).
Incident power-law is indicated by dotted line and is 
assumed to be isotropic. Reflected spectrum is emitted isotropically 
on the same side of the cloud, while 
outward spectrum is emitted  isotropically on the opposite side 
of the cloud. Transmitted spectrum is outward spectrum in 
normal direction to the observer.}
\label{fig:allspec}
\end{figure}

Generally our code models the  reflected and  outward  spectrum 
from the slab with assumption that both are emitted isotropically. 
In Fig.~\ref{fig:allspec} we present all spectra available 
for one particular cloud. The advantage of our code is that 
we do not separate absorbed and emitted spectra which 
are derived from two different parts of 
the solution of radiative transfer. This procedure is 
done in many previous attempts starting from Netzer (1993).
In {\sc titan} we solve transfer of absorption and emission continuum 
and lines  simultaneously producing outward spectrum (see 
Fig.~\ref{fig:allspec}), which is a mixture of absorption and 
emission features. Nevertheless, if a warm absorber cloud is on the 
line of sight to the observer, we see outward cloud spectrum 
emitted only in the direction normal to the slab, i.e. 
pointing to the observer. 

In this first paper we want to examine only the absorption features, 
therefore
the final spectra presented in the section below are 
purely absorption spectra, i.e., outward in the direction 
toward  the observer (transmitted). This is 
equivalent to the situation where the covering factor of 
the warm absorber is small, but a single warm absorber cloud is large and
located on the line of sight to the observer.
The model is applicable to Seyfert 1 galaxies since 
in those objects we observe mostly absorption lines. 
Although the absorbed spectra do not contain any emission lines, 
the temperature and ionization structure were computed with 
full transfer of absorbed, emitted and reflected radiation. 
Reflected and outward
spectra in other directions
will be treated more carefully in the next paper. 

In this paper generally we are discussing a grid of models with zero velocities
to see the main trends in the properties of 
the absorbing matter.
But we also present two models with non-zero turbulent velocity 
the same for all ions to trace how it affects
final equivalent widths of lines  
(see also a special case considered in \Agata et al. 2004).

\section{Grid of parameters and numerical limitation}
\label{sec:grid}

All computations are done assuming that the surface of 
a warm absorber has a number density $n_0=10^{11}$ cm$^{-3}$.
The exact value of this parameter seems to be unimportant. 
We have checked that adopting a lower value of the  surface density does not 
change the cloud structure.  
The ionization parameter $\xi$ gives a direct information
on the flux of the ionizing radiation which hits the surface of
the warm absorber. In our model we vary $\xi$ from $10^3$ up to $10^5$
erg cm s$^{-1}$. 

We assume that the ionizing continuum is coming from the center of
an active nucleus and has a power law shape.  
The photon index varies from 1.5 up to 2.5 as observed for the hard X-ray band.
The illuminating X-ray continuum in all models extends 
from 0.01 keV up to 100 keV.

When computing warm absorber models under constant pressure 
we deal with the well known problem of an illuminated
atmosphere with heavy elements. If the incident spectrum is hard
enough and/or the abundance of heavy elements is solar or higher,
a thermal instability develops (Fields 1965, Krolik, Mckee \& Tarter 1981)
which affects the conditions in the surface layers of the irradiated stellar
or accretion disk atmospheres, including AGN disks (e.g. R\'o\.za\'nska \& 
Czerny 1996). 

The instability occurs when for a single  value of the dynamical 
ionization parameter
$\Xi= F_{tot}/c P_{gas}$ the energy balance equation has three solutions
with different temperatures and densities.  Therefore, in a constant density
medium it is not possible to obtain any instability but they develop
if the density of the medium can adjust to the local conditions. Among these
three solutions, one  is unstable,
and two others are stable, but there is no stable solution which is continuous
through the whole irradiated layer.  
In our case we assume constant pressure instead of constant density and we
may expect thermal instabilities in the clouds with 
total column density high enough for the temperature to drop considerably 
inside the cloud
(see Sec.~\ref{sect:struc}).  

\begin{figure}
\epsfxsize=8.8cm \epsfbox[80 430 580 700]{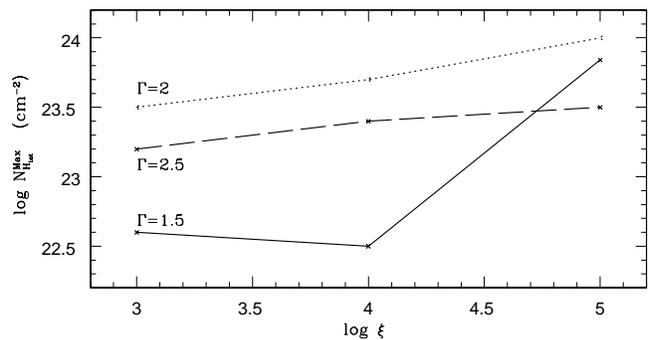}
\caption{The maximum total column density versus ionization parameter for 
each photon index. For a given value of $\Gamma$ and $\xi$ the structure of 
clouds with column densities higher than $ N^{Max}_{H_{tot}}$ cannot be 
determined because of thermal instabilities. }
\label{fig:grid}
\end{figure}

Generally, for each set of 
$\Gamma$ and $\xi$ we consider a range of values for the logarithm of the 
column density, $\log N_H$,
starting with 21. Whenever possible, we extended our grid to $\log N_H = 23.5$.
However, for specific values of the initial ionization parameter 
and/or steepness of X-ray spectrum 
we were not 
able to achieve thermal equilibrium. There is a maximum total column density,
$N^{Max}_{H_{tot}}$, for which instabilities are so strong that computations 
failed.  For this reason 
our grid of total column densities is not the same 
for each set of $\Gamma$ and $\xi$. These maximum values of the total
column density are shown in Fig.~\ref{fig:grid} for all combinations 
of $\Gamma$ and $\xi$.

The maximum value $N^{Max}_{H_{tot}}$ rises with an increase of the
ionization parameter but the relation is not strictly monotonic, and the
dependence on the spectral slope is even more complex. 
The most difficult case was $\Gamma=1.5$ and $\xi=10^4$, where
a temperature fall appeared even for low total column densities, and 
we were able to find a stable cloud only up to the 
$N_{H_{tot}}=3.16 \times 10^{22}$ cm$^{-2}$. Under those special conditions,
line cooling due to a single iron ion strongly 
dominated all other cooling mechanisms, and the medium was optically thick
in the emitted lines. Therefore, the local temperature strongly depended on
the presence or absence of a layer {\it behind} the considered zone. The 
cooling of the outer zone served as a heating mechanism 
to the inner zone, compensating exactly the losses. As a result,
the Thomson depth of the hot zone depended roughly linearly on the assumed
$N_{H_{tot}}$.

In this paper we also study the case when absorbing matter has turbulent 
velocity. For this purpose we consider two cases, one with $v_{turb}=100$ km/s,
and one with $v_{turb}=300$ km/s, the same for each ion.
 
\section{Structure of the warm absorber}
\label{sect:struc}

\begin{figure}
\epsfxsize=8.8cm \epsfbox[100 220 580 700]{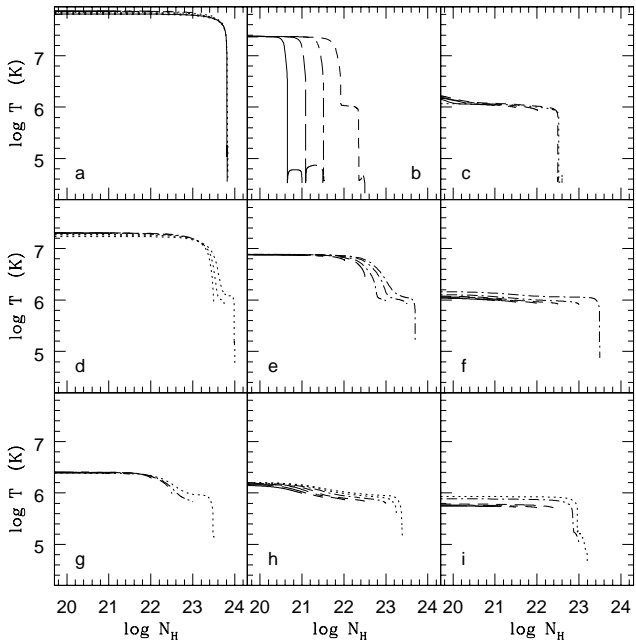}
\caption{The temperature profiles within the irradiated clouds at constant 
pressure. The different panels show the dependence on the total column density, 
starting with $\log N_{H_{tot}} = 21$, with step sizes of 0.5. The step
size decreases to 0.1 when approaching the value of $\log N^{Max}_{H_{tot}}$.
Panels a, b, c  show the results for $ \xi =10^5$, 
$10^4$ and $10^3$, and for $\Gamma =1.5$. Panels d, e ,f show respectively the
same ionization parameters but for  $\Gamma =2$. 
The last three panels, g, h, i,  present the cases for $\Gamma =2.5$.}
\label{fig:temp}
\end{figure}

We expect that at the illuminated
face of a cloud the temperature will be high and density low, 
and a high temperature
equilibrium on the ``S curve'' (Krolik, Mckee \& Tarter 1981) will be
reached. 
This layer is expected to be  optically thin despite some lines which 
may be saturated. 
Going deeper inside the warm absorber the illuminating continuum will be
absorbed by heavy elements and we expect that the temperature structure 
will show a strong decrease with accompanying increase of the density, 
while the ionization state decreases, 
but not so rapidly. Therefore after this geometrically extended optically 
thin layer, we will get geometrically thin layers with strong 
temperature and density gradients becoming optically thicker. 
 
\begin{figure}
\epsfxsize=8.8cm \epsfbox[100 220 580 700]{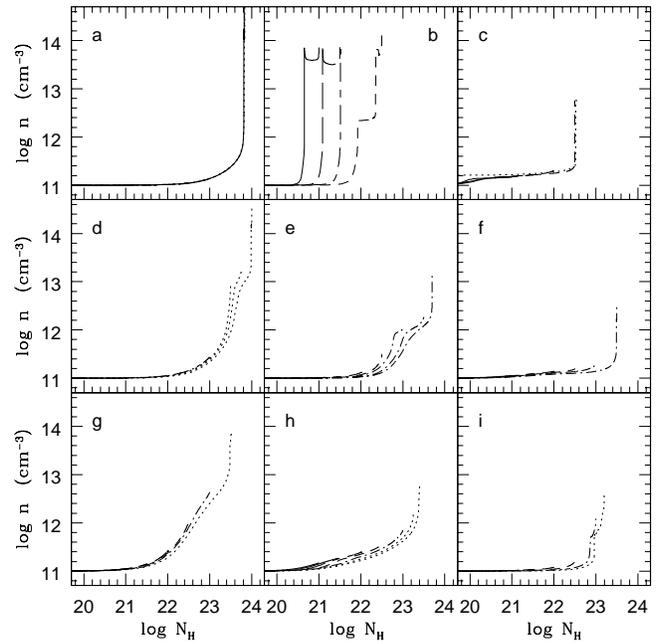}
\caption{The density profiles within the irradiated clouds at constant 
pressure. The different panels show the dependence 
on the total column density, 
starting with $\log N_{H_{tot}} = 21$, with step sizes of 0.5. The step
size decreases to 0.1 when approaching the value of $\log N^{Max}_{H_{tot}}$.
Panels a, b, c  show the results for $ \xi =10^5$, 
$10^4$ and $10^3$, and for $\Gamma =1.5$. Panels d, e, f show respectively the
same ionization parameters but for  $\Gamma =2$. 
The last three panels, g, h, i,  present the cases for  $\Gamma =2.5$.}
\label{fig:gest}
\end{figure}

\begin{figure}
\epsfxsize=8.8cm \epsfbox[100 350 580 700]{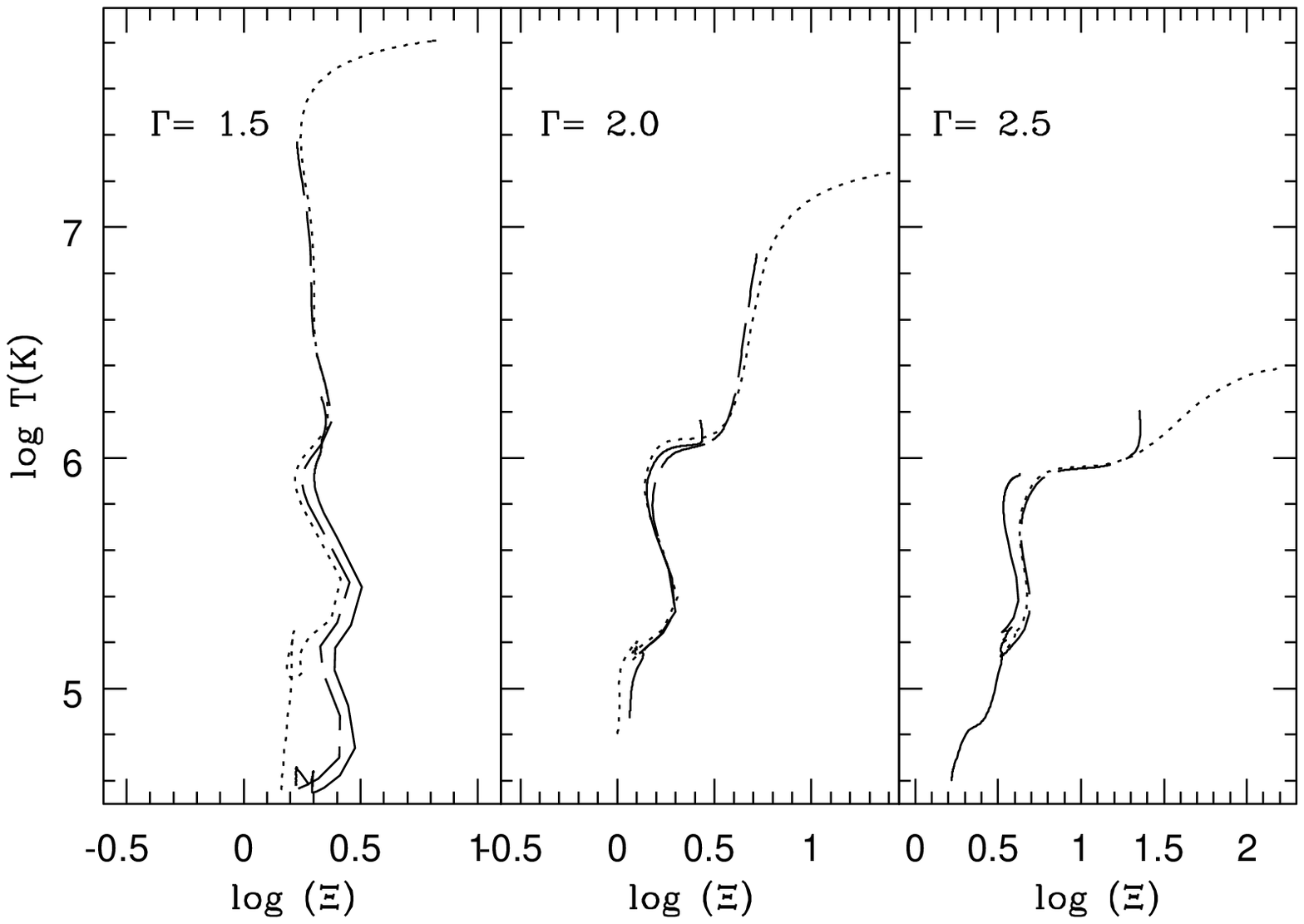}
\caption{The thermal equilibrium curves for the clouds at constant 
pressure. Clouds 
with $\Gamma=1.5$ are presented in the left panel, with $\Gamma=2.0$ in the
middle panel,
and with $\Gamma=2.5$ in the right panel. Lines have following meaning:
 dotted line - $\xi=10^5$, dashed  line - $\xi=10^4$,
and solid line for clouds with $\xi=10^2$.
In all cases we show clouds with maximum total column density.  }
\label{fig:xibig}
\end{figure}

The temperature structures of all computed  clouds are 
given  in Fig.~\ref{fig:temp}. Models with 
$\Gamma=1.5$ are presented in panels a, b, c for 
$\xi= 10^5, 10^4 $ and $10^3$  respectively. 
The same sequence of $\xi$,  but for $\Gamma=2$, is shown 
in panels d, e, f, and the softest intrinsic spectrum 
with photon index 2.5 is presented in g, h and i panels. 
The density structure for exactly the same 
set of models is presented in Fig.~\ref{fig:gest}. 

For each set of $\Gamma$ and $\xi$ we present results  
for a grid of column densities starting from $N_{H_{tot}} =10^{21}$
up to about $N^{Max}_{H_{tot}}$, depending on the appearance of
thermal instabilities. We use a 0.5 step size in 
logarithm of the total column density. 
  
In each panel (except for b, where $\Gamma=1.5$ and $\xi=10^4$) 
we see the same trend in the structure while increasing 
the value of the total column density. 
For the lowest total column densities  the temperature does not decrease
strongly, because the cloud is thin enough to exist only
in the high temperature equilibrium. In such a case the warm absorber is 
well modeled by a constant density slab with low density 
of the order of $n=10^{11}$ cm$^{-3}$.
A temperature drop appears when $N_{H_{tot}}$ increases,
and the highly ionized hot layer is complemented by geometrically thin, 
dense zones reaching even $n=5 \times 10^{14}$
cm$^{-3}$.
Since the transition to solutions with the outer cool cloud layer 
occurs rapidly, we had to decrease the step
in $ \log N_{H_{tot}}$ to 0.1, for the last two curves in 
almost each panel in order to resolve correctly the structure.  

The exceptional case of $\Gamma=1.5$, $\xi=10^4$ 
(panel b in Fig.~\ref{fig:temp}) which was discussed at the end of 
Section~\ref{sec:grid} shows a strong temperature jump even for 
$N_{H_{tot}} =10^{21}$ cm$^{-2}$. The dense layer on the back of the 
illuminated
face has an unphysical temperature bump. Such a layer has 
a non negligible optical thickness and it is seen in modeled spectra, which
possesses strong absorption (see Section below).

In Fig.~\ref{fig:xibig} we present thermal equilibrium curves, 
$T(\Xi)$, for several 
clouds. The value of dynamical ionization parameter 
across the cloud was computed using relation 
$\Xi=P_{rad}/P_{gas}$.
From this figure it is clearly seen the difference of our 
models with those when two constant density slabs are used to fit 
a data. We propose warm absorber as a single cloud which passes 
through all ionization states, and each layer interacts radiatively
with another one as it is usually done in classical atmospheric calculations. 

\section{Modeled spectra}
\label{sect:local_spot}

\begin{figure}
\epsfxsize=8.8cm \epsfbox[80 220 580 700]{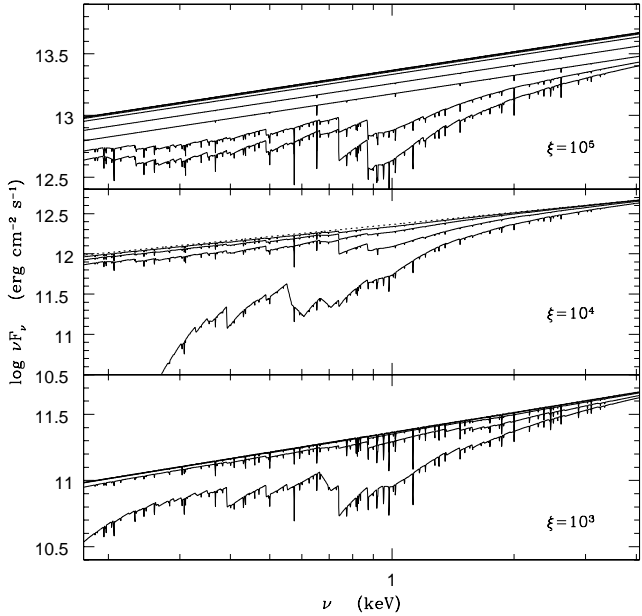}
\caption{Set of transmission spectra for the photon index of the incident
continuum $\Gamma = 1.5$ and $\xi = 10^5$ (upper panel), $\xi = 10^4$ (middle
panel) and $\xi = 10^3$ (lower panel). The results for a 
range of $\log N_{H_{tot}}$
are shown in each panel starting from $\log N_H=21.$ up to 
$\log N^{Max}_{H_{tot}}$.}
\label{fig:spec05}
\end{figure}

\begin{figure}
\epsfxsize=8.8cm \epsfbox[80 220 580 700]{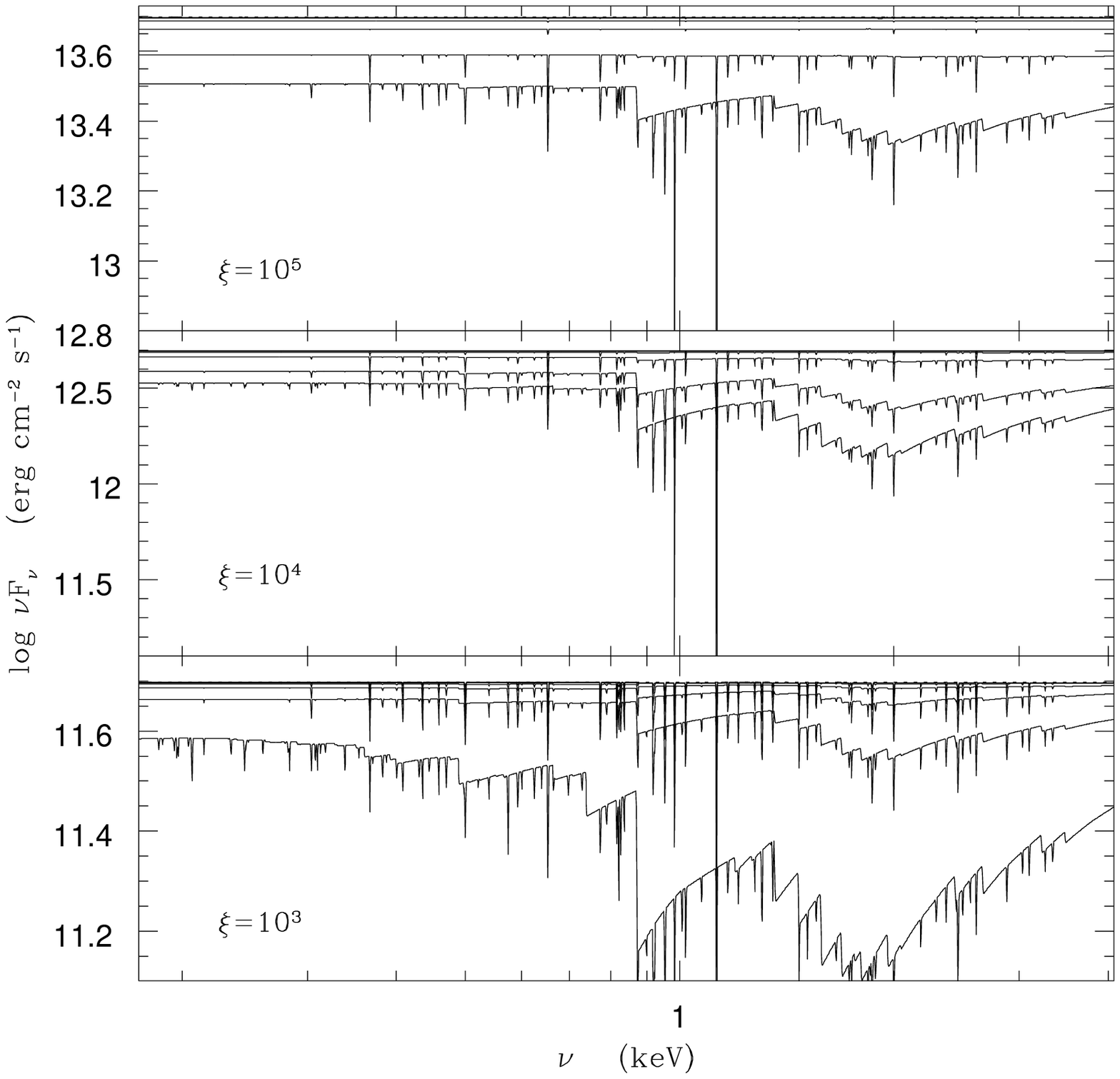}
\caption{Set of transmission spectra for the photon index of the incident
continuum  $\Gamma = 2.0$ and $\xi = 10^5$ (upper panel), $\xi = 10^4$ (middle
panel) and $\xi = 10^3$ (lower panel). The results for a range of
 $\log N_{H_{tot}}$
are shown in each panel starting from $\log N_H=21.$ up to 
$\log N^{Max}_{H_{tot}}$.}
\label{fig:spec10}
\end{figure}

\begin{figure}
\epsfxsize=8.8cm \epsfbox[80 220 580 700]{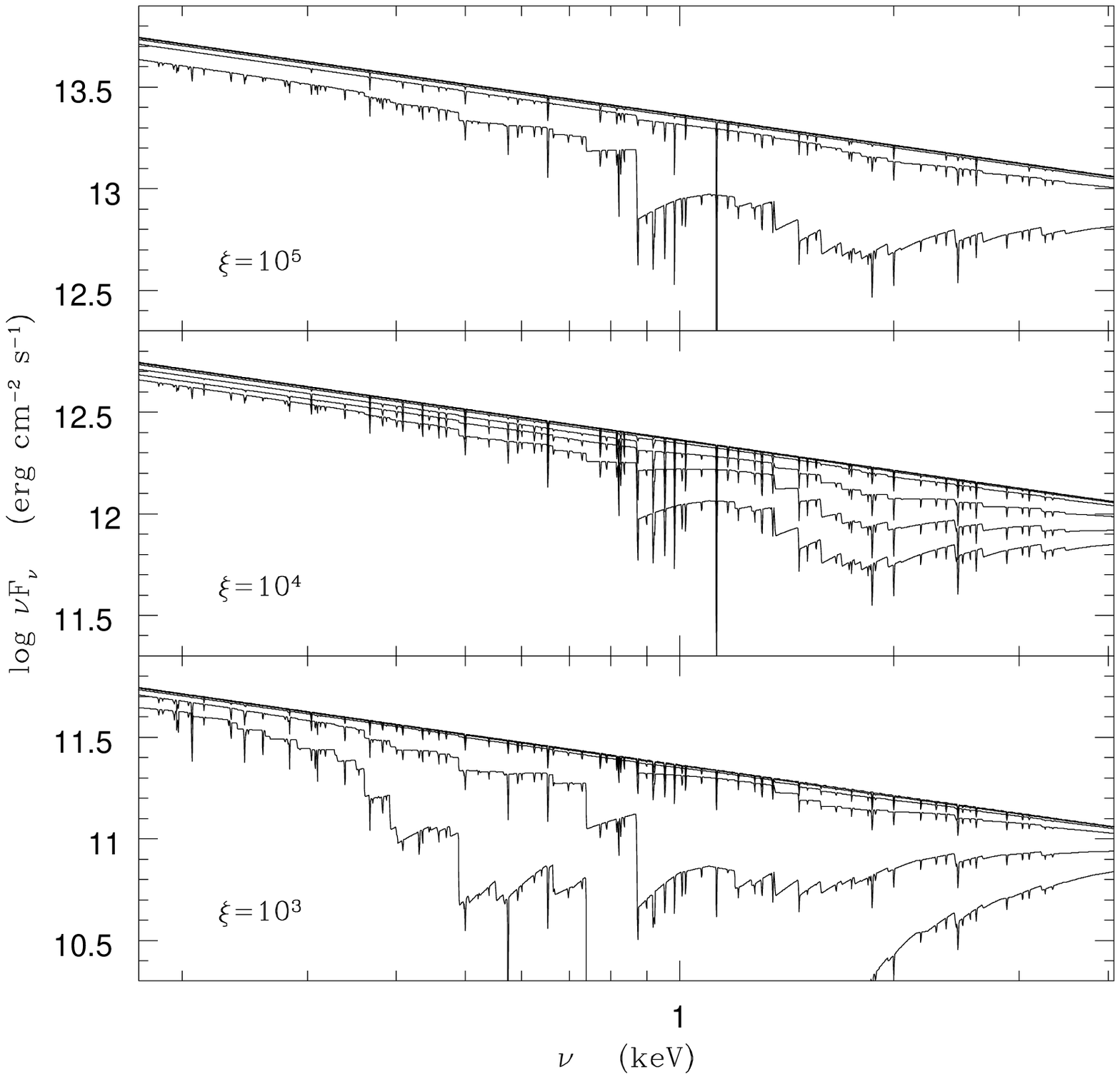}
\caption{Set of transmission spectra for the photon index of the incident
continuum $\Gamma = 2.5$ and $\xi = 10^5$ (upper panel), $\xi = 10^4$ (middle
panel) and $\xi = 10^3$ (lower panel).  The results for a range 
of $\log N_{H_{tot}}$
are shown in each panel starting from $\log N_H=21.$ up to 
$\log N^{Max}_{H_{tot}}$.}
\label{fig:spec15}
\end{figure}

The transmitted spectra for the whole set of models are presented in 
Figs.~\ref{fig:spec05}, \ref{fig:spec10}, and \ref{fig:spec15}
for $\Gamma=1.5, 2 $ and 2.5 respectively. 
All figures are given in the range of energies  between 0.1 and 4 keV, where
most lines are present. The energy ranges of iron lines from 6 keV up to 8 keV 
are presented and discussed in Sec.~\ref{sec:rel}.

For high ionization parameters and low total column densities 
the spectra are featureless since
the matter is almost fully ionized and only changes in the  continuum are caused 
by Comptonization. This effect is clearly seen 
in the upper panels of Figs.~\ref{fig:spec05} and \ref{fig:spec10}, 
where the whole outgoing continuum is lowered. 
The Compton heating reaches almost 50\% of total heating,
therefore the
determination of the  exact value of Compton  heating
was very important in such cases and we did it iteratively using 
the Monte Carlo code {\sc noar}.    

In the spectra corresponding to higher total column densities 
we observe
many absorption features. The spectrum in the energy range 
from 0.8 keV up to 2 keV 
is especially strongly absorbed due to many transitions of highly ionized heavy 
elements.    

\begin{figure}
\epsfxsize=8.8cm \epsfbox[80 220 580 700]{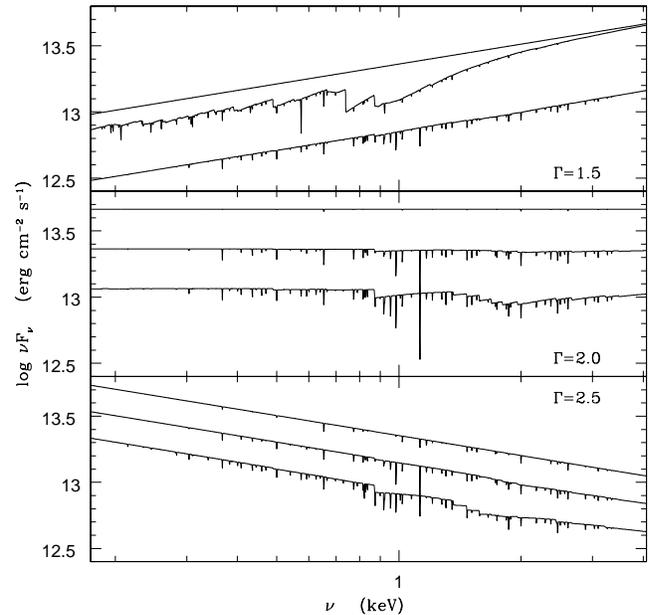}
\caption{Comparison of spectra for the same $N_{H_{tot}}$, but 
for different ionization parameters. In upper panel we compare 
clouds of $N_{H_{tot}}= 10^{22}$ cm$^{-2}$ illuminated by 
radiation with photon index  $\Gamma=1.5$, middle panel 
$N_{H_{tot}}= 10^{23}$ cm$^{-2}$ and $\Gamma=2.0$, and lower panel
$N_{H_{tot}}= 3.16\times 10^{22}$ and $\Gamma=2.5$.
In each panel lines from the top to the bottom present cases for  
$\xi=10^5, 10^4$ and $10^3$ respectively.}
\label{fig:speion}
\end{figure}

The comparison of spectra for the same total column density of the cloud
but for different ionization parameters is shown in Fig.~\ref{fig:speion}.
The upper panel presents the results for slabs 
with $N_{H_{tot}}= 10^{22}$ cm$^{-2}$ illuminated
by radiation of photon index $\Gamma=1.5$. 
The different spectra from the top to the bottom 
represent the cases for $\xi=10^5, 10^4$ and $10^3$ respectively.
The same sequence of ionization parameters is presented
in the middle panel but for $N_{H_{tot}}= 10^{23}$ cm$^{-2}$ and $\Gamma=2.0$,
and in the lower panel but for $N_{H_{tot}}= 3.16\times 10^{22}$  
cm$^{-2}$ and $\Gamma=2.5$.

\begin{figure}
\epsfxsize=8.8cm \epsfbox[100 220 580 700]{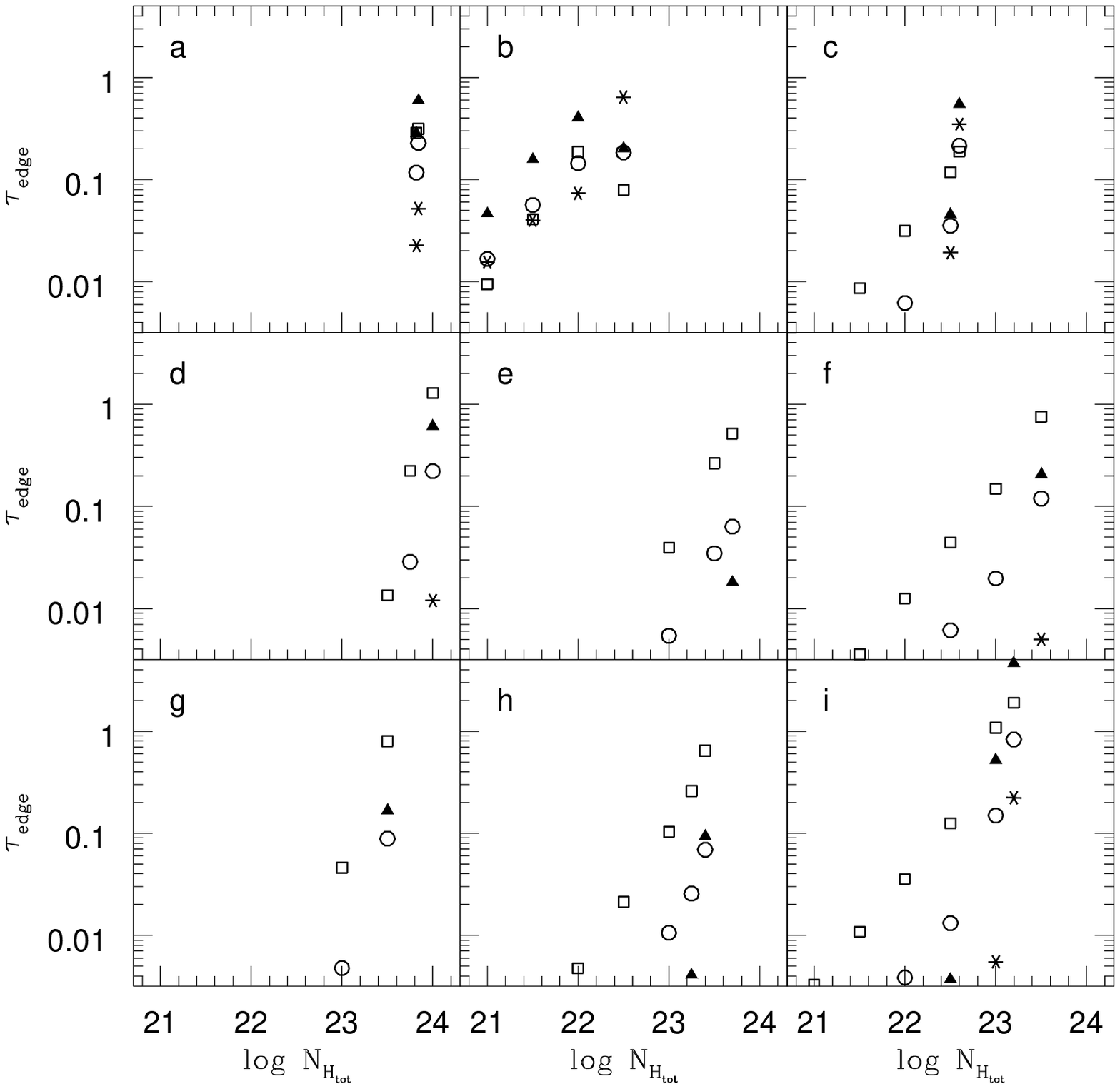}
\caption{The strength of absorption edges for irradiated clouds at constant 
pressure. The different panels show the dependence on the total
 column density, 
starting with $\log N_{H_{tot}} = 21$, with step size of 0.5. The step
size decreases to 0.1 when approaching the value $\log N^{Max}_{H_{tot}}$.
Panels a, b, c  show the results for $ \xi =10^5$, 
$10^4$ and $10^3$, and for $\Gamma =1.5$. Panels d, e, f show respectively the
same ionization parameters but for  $\Gamma =2$. 
Last three panels: g, h, i  present the cases for  $\Gamma =2.5$.
Filled triangles denote  O{\sc vii} edges, 
open squares are O{\sc viii}, stars are C{\sc v}, and 
open circles are C{\sc vi} edges.}
\label{fig:prog}
\end{figure}

The spectrum in the case of  $\Gamma=1.5$ and  $\xi=10^4$ is again 
exceptional because of the numerical problems mentioned above. 
In this spectrum many absorption edges are
seen. This is because the optically thick layer on the back of illuminated cloud
is also geometrically thick and the absorption is much stronger. 
Additionally, the highly ionized optically thin layer on the front of illuminated
cloud  is thin enough to not interact with X-ray radiation at all, explaining 
the lack of lines above 1 keV in this spectrum. 

Other spectra show an increase in the number of lines with 
decreasing ionization parameter. 
Also less energetic lines (below 1 keV) are much more numerous for 
lower $\xi$. This is because the back 
side of an illuminated cloud reaches lower temperatures and higher 
densities, and also lower ionization states for lower $\xi$.

\subsection{Absorption edges}

The detection of absorption edges in the spectra of AGN puts direct constraints
to the total column density of the warm absorber. Therefore, we show 
in Fig.~\ref{fig:prog} the strength of the main ionization edges 
in our synthetic spectra 
for comparison with observations. Models with 
$\Gamma=1.5$ are presented in panels a, b, c for 
$\xi= 10^5, 10^4 $ and $10^3$  respectively. 
The same sequence of $\xi$,  but for $\Gamma=2$ is shown 
in panels d, e, f, and the softest intrinsic spectrum 
with photon index 2.5 is presented in g, h and i panels. 

The strength of an edge is measured by its total optical thickness defined
by total absorption plus scattering opacity coefficient multiplied by the density
and integrated over the geometrical size of cloud.
We show only O{\sc vii} (filled triangles), O{\sc viii} (open squares),
C{\sc v} (stars), and C{\sc vi} (open circles) edges above 
$\tau_{edge} = 0.003$,
which can be resolved using presently working X-ray satellites. 

Again panel b shows the exceptional case, and we see that the ionization
edges are strong for the reason 
mentioned in the previous Section. 
In the other cases the edges become visible  only when the
 total column 
densities are higher than 10$^{23}$ cm$^{-2}$ for $\xi=10^5$ (panels a, d, g). 
This minimum total column density decreases with decreasing ionization 
parameter
reaching the value of 10$^{22}$  cm$^{-2}$  for $\xi=10^3$.  

\subsection{Strength of the absorption lines}

\begin{figure}
\epsfxsize=8.8cm \epsfbox[80 220 580 700]{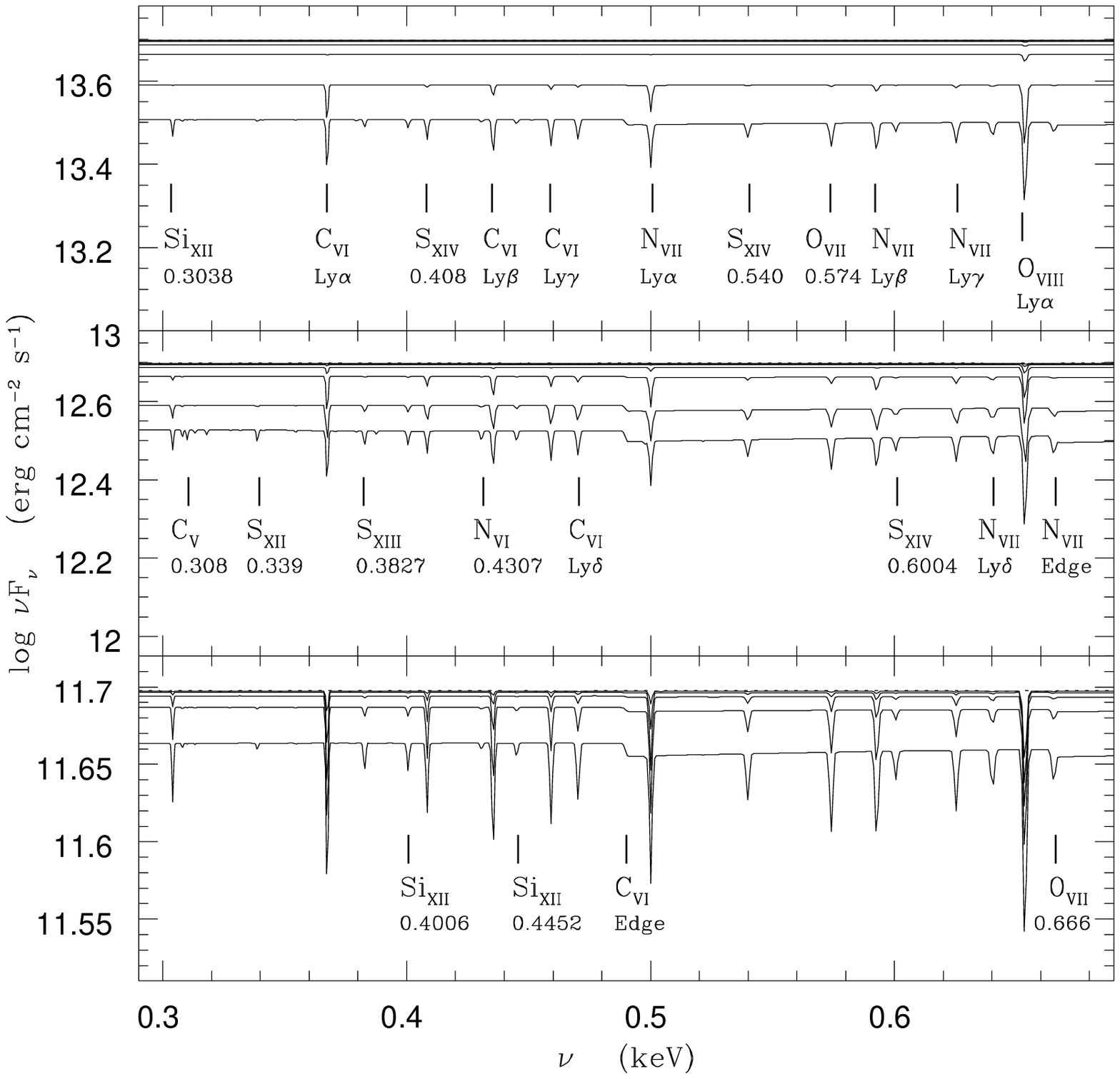}
\caption{The expanded version of Fig.~\ref{fig:spec10} ($\Gamma = 2$) in the 
softest energy range between 0.3 and 0.7 keV. 
Models for different $\xi=10^5, 10^4$,
and $10^3$ are presented in upper, middle and lower panel
respectively. The lowest continuous lines in each
panel correspond to $N^{Max}_{H_{tot}}$ appropriate for the set
of $\Gamma$ and $\xi$ as seen if Fig.~\ref{fig:grid}.}
\label{fig:li10}
\end{figure}

\begin{figure}
\epsfxsize=8.8cm \epsfbox[80 220 580 700]{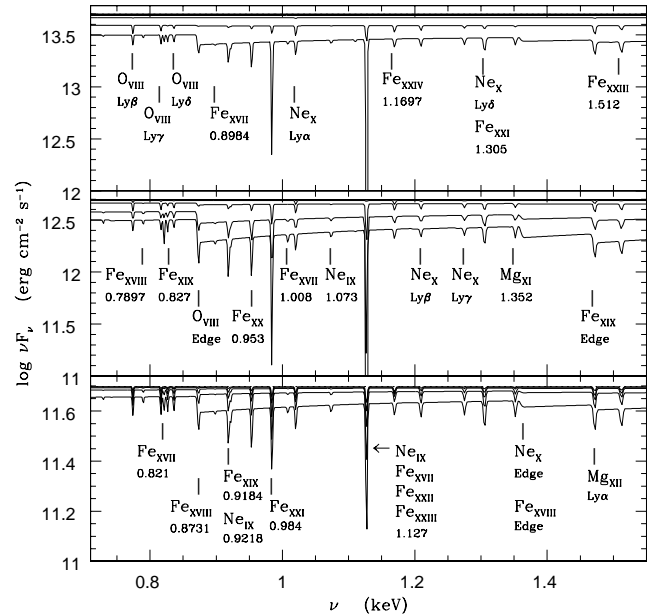}
\caption{The same as Fig.~\ref{fig:li10} but for energy range between 0.7
and  1.5 keV.}
\label{fig:mid10}
\end{figure}

\begin{figure}
\epsfxsize=8.8cm \epsfbox[80 220 580 700]{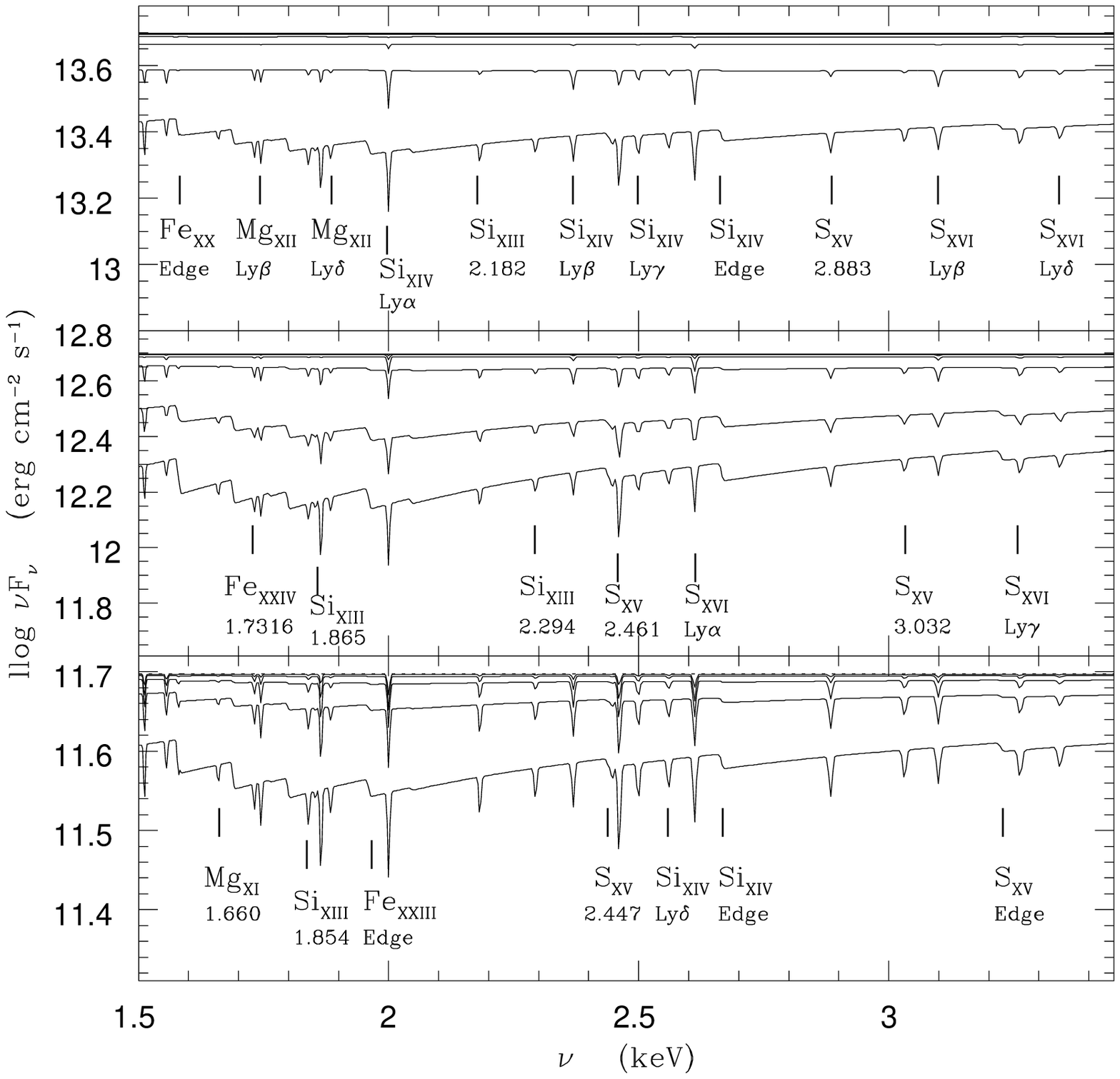}
\caption{The same as Fig.~\ref{fig:li10} but for energy range between 1.5
and  3.5 keV.}
\label{fig:hard10}
\end{figure}


\begin{figure}
\epsfxsize=8.8cm \epsfbox[80 220 580 700]{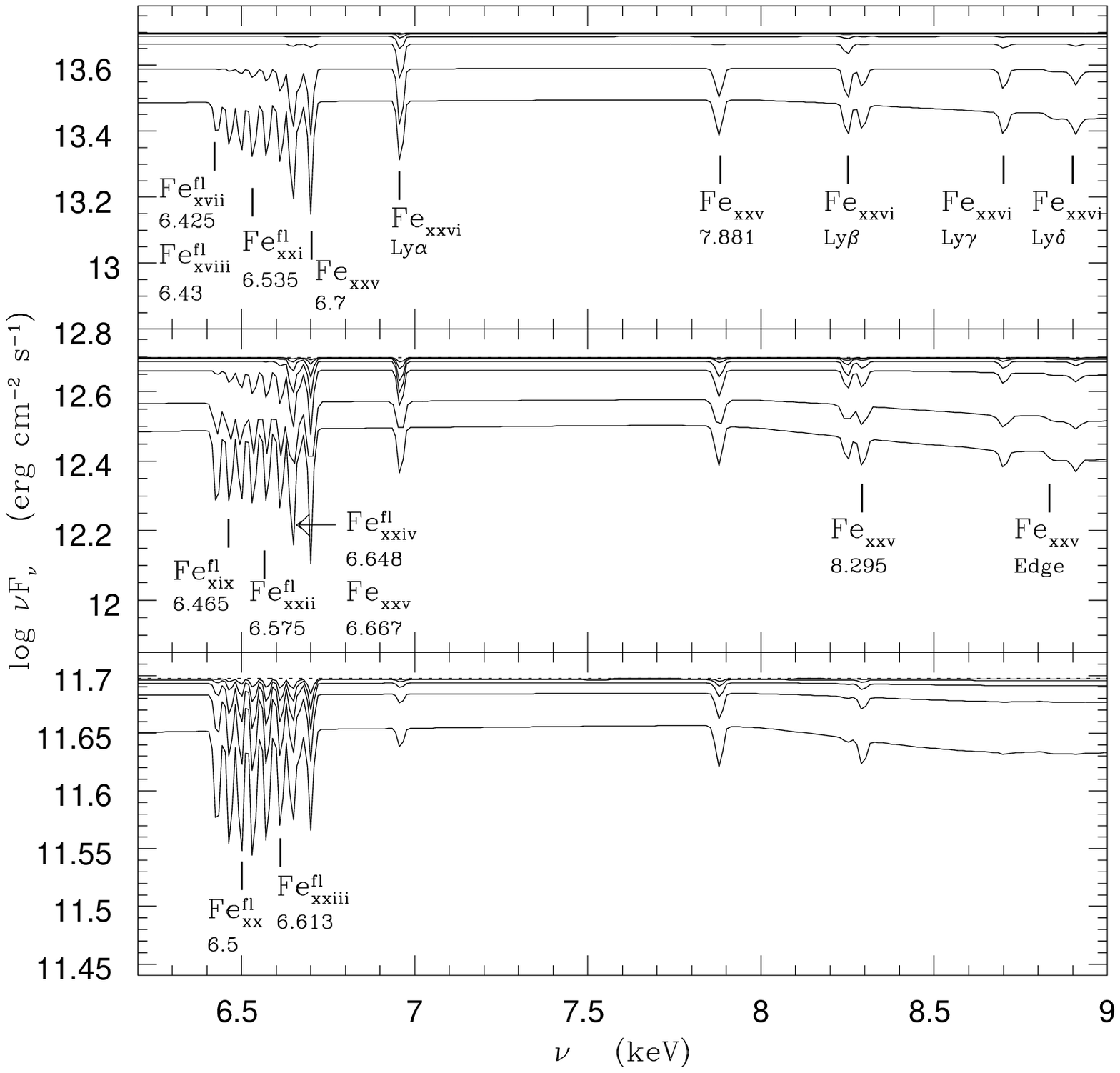}
\caption{The same as Fig.~\ref{fig:li10} but for energy range between 6.2
and  9 keV.}
\label{fig:fe10}
\end{figure}


\begin{table*}

\caption{Equivalent widths in [eV] 
of absorption lines from  clouds  with  different 
ionization parameters at the surface.   \label{tab:linie1}}
\begin{center}     
\renewcommand{\arraystretch}{1.1}
\begin{tabular}{|c|r|r|r|r|c|c|l|l|}  
\hline
$ \Gamma$  & \multicolumn{3}{|c|}{2.0}&   2.5 & 2.0 & 2.0 & Observed & ref. \\
\hline 
$\xi$  & {$10^5$} & {$10^4$} & 
 {$10^3$} &{$10^3$} & {$10^4$} & {$10^4$} &  &  \\
\hline
$\log (N_{H_{tot}1})$ &  23.5 & 23.5   & 23.5 & 23.2 &  23.5 &  23.&  & \\
                      &       &   & & & $v_{turb}$ =100 km/s 
& $v_{turb}$ =300 km/s & & \\
\hline 
\hline 
C{\sc vi} $Ly {\alpha}$  & 0.18 & 0.24 &  0.25 &  0.31 & 0.5 &  1.27 &   
 $0.98\pm 0.33$&  (2)  \\
\hline 
N{\sc vii} $ Ly {\alpha}$ & 0.19& 0.28 & 0.31 & 0.4 & 0.63 & 1.55 & 
 $1.33\pm 0.24$ & (2) \\
\hline 
O{\sc viii} $Ly {\alpha}$ & 0.54 & 0.65 & 0.79 & 0.98 & 1.13 & 2.97&  
 $1.85 \pm 0.55$& (1) \\
\hline 
O{\sc viii} (0.7743)  & 0.38 & 0.44& 0.44& 0.38 & 1.05 & 2.76 & 
 $1.77 \pm 0.44$ & (1) \\
\hline
Fe{\sc xvii}  (0.8211) & 0.002& 0.35 & 0.93 & 0.90 & 0.7 & 1.24 &  
 $1.42 \pm 0.28$ &  (1)\\
\hline
Fe{\sc xix} (0.8266) &0.002 & 0.33  & 0.52 & 0.47 & 0.97 & 2.50 &  
$0.83 \pm 0.24$ & (1) \\ 
\hline 
Fe{\sc xviii} (0.8731)  & 0.01  & 0.45 & 0.88 & 0.47 & 1.0 & 2.37 &  
$1.52 \pm 0.34$&  (1) \\
\hline
Fe{\sc xvii} (0.8984) &-- & 0.1  & 0.23 & 0.29 & 0.2 & 0.2 &  
$1.43 \pm 0.52$ &  (2) \\
\hline
Fe{\sc xix} (0.9184) & 0.1& 0.96  & 1.17 & 0.96 & 1.47  & 3.48  & 
 $1.84 \pm 0.54$  blend & (2)  \\ 
\hline
Ne{\sc ix} (0.9218) &0.03 & 0.33 & 0.63  & 1.25 &  0.8 & 1.48  & 
   $4.13 \pm 0.48$  & (1) \\
 & & & & &  & &  $2.6 \pm 1.48$ & (2)  \\
 & & & & &  & & $1.33^{+0}_{-0.30}$ & (3) \\
 & & & & &  & &  $1.09^{+0.35}_{-0.34}$ & (4) \\
\hline
Fe{\sc xx} (0.953) & 0.2 & 1.22  & 1.22 & 0.97 & 1.69 & 3.95 & 
$4.88 \pm 0.42$  blend &  (1) \\ 
\hline
Ne{\sc x}  $Ly {\alpha}$ & 0.6 & 0.77 & 0.86 & 0.78 & 1.54 &4.08 & 
$3.20 \pm 0.24$  blend & (1) \\
\hline
Ne{\sc ix}    (1.127)  & 0.003 & 0.13  & 0.27& 0.27 & 0.2 &0.2  & all complex    &   \\ 
Fe{\sc xvii}  (1.127)  & -- & 0.04  & 0.2  & 0.22 & 0.07 &0.07 &
$4.66 \pm 0.33$ &(1)  \\ 
Fe{\sc xxiii} (1.127)  & 0.964 & 3.82 & 2.86 & 1.12 & 4.54  &9.55  &
 $2.15 \pm 0.82$&  (2) \\ 
\hline
Ne{\sc x}  (1.209) & 0.36 & 0.54  & 0.55 & 0.46 & 1.40  &3.42  &
$2.3 \pm 0.24 $ & (1) \\  
\hline 
Mg{\sc xii}  $Ly {\alpha}$  & 0.74 & 0.99 & 1.08 & 0.82 & 2.09 &5.45 &  
$4.39 \pm 0.24$ & (1)  \\ 
\hline                  
Fe{\sc xxiv} (1.556)  & 0.43 & 0.5 & 0.44 & 0.1 &  1.69 &3.94 &  
 $1.84 \pm 0.23$&  (1) \\ 
\hline
Mg{\sc xii}  (1.745)  & 0.4 & 0.67 & 0.7 & 0.56 & 1.83 &4.09 &  
  $1.75 \pm 0.22$ & (1) \\ 
\hline
Si{\sc xiii} (1.865)  & 0.46 & 1.47 & 1.92 & 1.91 & 2.66 &6.6 &  
 $4.18 \pm 0.20$ & (1)\\
 & & & & & & &  $3.37 \pm 1.68$ & (2) \\
 & & & & & & &  $1.49^{+0.76}_{-0.73} $ & (4)\\
\hline
Si{\sc xiv}  $Ly {\alpha}$ & 1.26 & 1.78 & 1.80 & 1.13 & 3.12  & 7.99 & 
$6.61 \pm 0.26$ & (1)  \\
 &  & & & & & & $3.87 \pm 1.94$&  (2) \\
 & & & & & & & $4.19^{+1.42}_{-1.29} $ & (4) \\
\hline 
Si{\sc  xiv} (2.370) & 0.8 & 0.97 & 0.94 &  0.71 & 2.72 &6.64 & 
$5.03 \pm 0.72$ blend &  (1)  \\ 
\hline
S{\sc xv} (2.461)  & 0.72 & 2.23 & 2.56 & 2.02 & 3.62  &8.73 & 
 $4.49 \pm 0.59$ &  (1)  \\ 
\hline
Si{\sc  xiv} (2.500)  & 0.48 & 0.84 & 0.85 &  0.63 & 2.33  &4.48 & 
$1.96 \pm 0.71$ &  (1)  \\ 
\hline
S{\sc xvi}  $Ly {\alpha}$  (2.611)  & 1.51 & 1.93 & 1.81 & 0.96 & 3.68 &9.3 & 
 $5.88 \pm 0.66$ &  (1)\\
 &  & & & & & &  $3.85^{+2.86}_{-2.80} $ & (4) \\
\hline
S{\sc xv} (2.883)   & 0.3 & 0.93 & 1.04 & 0.86 & 2.87  &6.15 & 
$1.81 \pm 0.67$ & (1)  \\ 
\hline
S{\sc xv} (3.032)   &0.13  &  & 0.9 & 0.72 & 2.19  &3.31 & 
$2.82 \pm 0.82$ & (1) \\
\hline 
S{\sc xvi}  (3.100)  & 0.9 & 1.11 & 1.04 & 0.69 & 3.10 &6.48 & 
$2.87 \pm 0.78$ blend  & (1)  \\ 
\hline
Fe{\sc xxv} (6.700)  & 6.74 & 9.0 & 5.35 & 0.25 & 11.99 &25.19 &  
 $13.76 \pm 5.07$ & (1) \\ 
\hline
Fe{\sc xxv} (7.881)  &  3.37 & 3.57 & 2.22 & 0.047 & 8.25 &17.75 & 
$33.06 \pm 19.04$ & (1)  \\ 
\hline     

\end{tabular} 
\end{center}
(1) NGC 3783, 900 ksec {\it Chandra} Kaspi et al. 2002; 
(2) NGC 5548, 86.4 ksec  {\it Chandra} Kaastra et al. 2002;
(3) Mrk 509, 59 ksec {\it Chandra} Yaqoob et al. 2003;
(4) NGC 4051, 81.5 ksec {\it Chandra} Collinge et al. 2001. 
\end{table*}

Several physical conditions influence the amount of energy absorbed in 
lines: the heavy element abundance, temperature of the medium, the hardness of 
the illuminating continuum, the column densities of particular ions, 
and the turbulent
velocity of the absorbing matter. 
It is not obvious which condition is the most important. 
The curve of growth analysis should be done, but at present 
 in our basic models we didn't
allow for any velocity dispersion in the absorbing gas.

In order to study the strength of absorption lines in our spectra,
in Figures \ref{fig:li10}, \ref{fig:mid10}, \ref{fig:hard10},
and \ref{fig:fe10} we show expanded views of the spectra 
corresponding to $\Gamma=2$ 
(see Fig.~\ref{fig:spec10}).
The range between 0.3 and 0.7 keV is shown in Fig.~\ref{fig:li10}, 
between 0.7 and 1.5 keV in Fig.~\ref{fig:mid10}, from 1.5 up to 3.5 keV
is presented in Fig.~\ref{fig:hard10}, and finally the range from 6.2 up to 9 keV
is shown in Fig~\ref{fig:fe10}. 

All lines are weak (see Table~\ref{tab:linie1}), with equivalent 
widths smaller than 1 eV, and only in a few cases (Si{\sc xiv}  $Ly {\alpha}$,
S{\sc xvi}  $Ly {\alpha}$, Fe{\sc xxvi} $Ly {\alpha}$ up to $Ly {\gamma}$,
also some  fluorescent iron lines) 
do they reach a few eVs. For a decreasing ionization parameter $\xi$ 
(going from upper to bottom panels on each figure) we see that the lines
become stronger and are present even for lower total column densities. 

Some lines do not change very much with increasing column densities. 
These lines are predominantly from the most highly ionized species. 
Lines from ions of slightly lower ionization
degree are not present in cases when the total column density is low, because
the dense zone does not exist on the back of illuminated cloud. 

Several modeled lines do agree with those detected by 
X-ray satellites as is shown in Tab.~\ref{tab:linie1}.
But most of them are slightly lower which may 
indicate that the observed lines are blended. 
We do expect this effect with present resolution of X-ray gratings.
Also, it may indicate that material is in
turbulent motion. 

\begin{figure}
\epsfxsize=8.8cm \epsfbox[20 350 530 700]{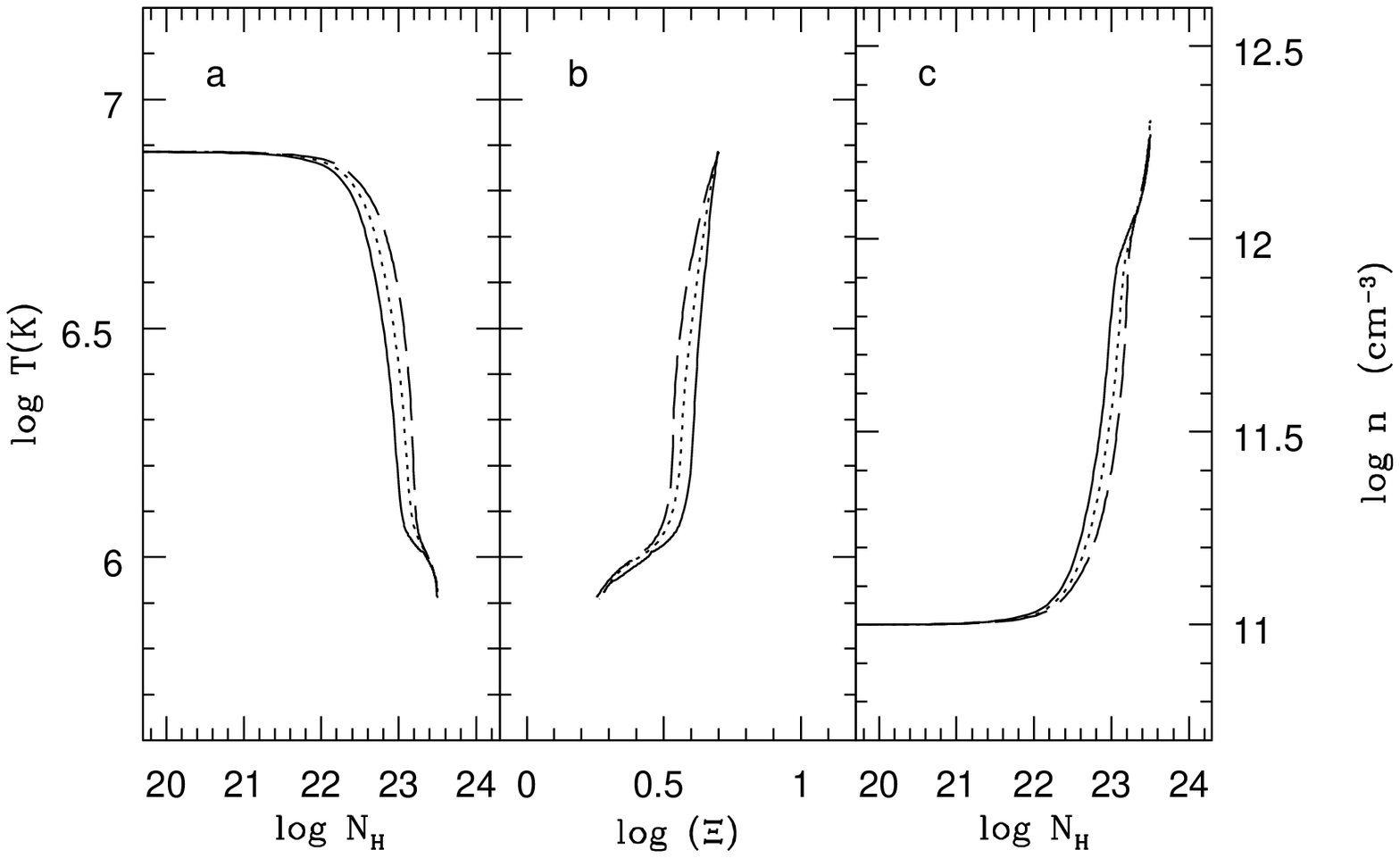}
\caption{The structure of warm absorber in case
of turbulent velocity for  $\Gamma = 2.0$, $\xi = 10^4$ 
and maximum column density $N_{Htot}^{Max}$. 
Panel a) shows temperature structure,
b) thermal equilibrium curve $T(\Xi)$ and 
panel c) density structure. 
Solid line represents the case with $v_{turb}=0$, dotted line
with $v_{turb}=100$ km/s, and dashed line with  $v_{turb}=300$ km/s .}
\label{fig:turb1}
\end{figure}

\begin{figure}
\epsfxsize=8.8cm \epsfbox[80 190 580 700]{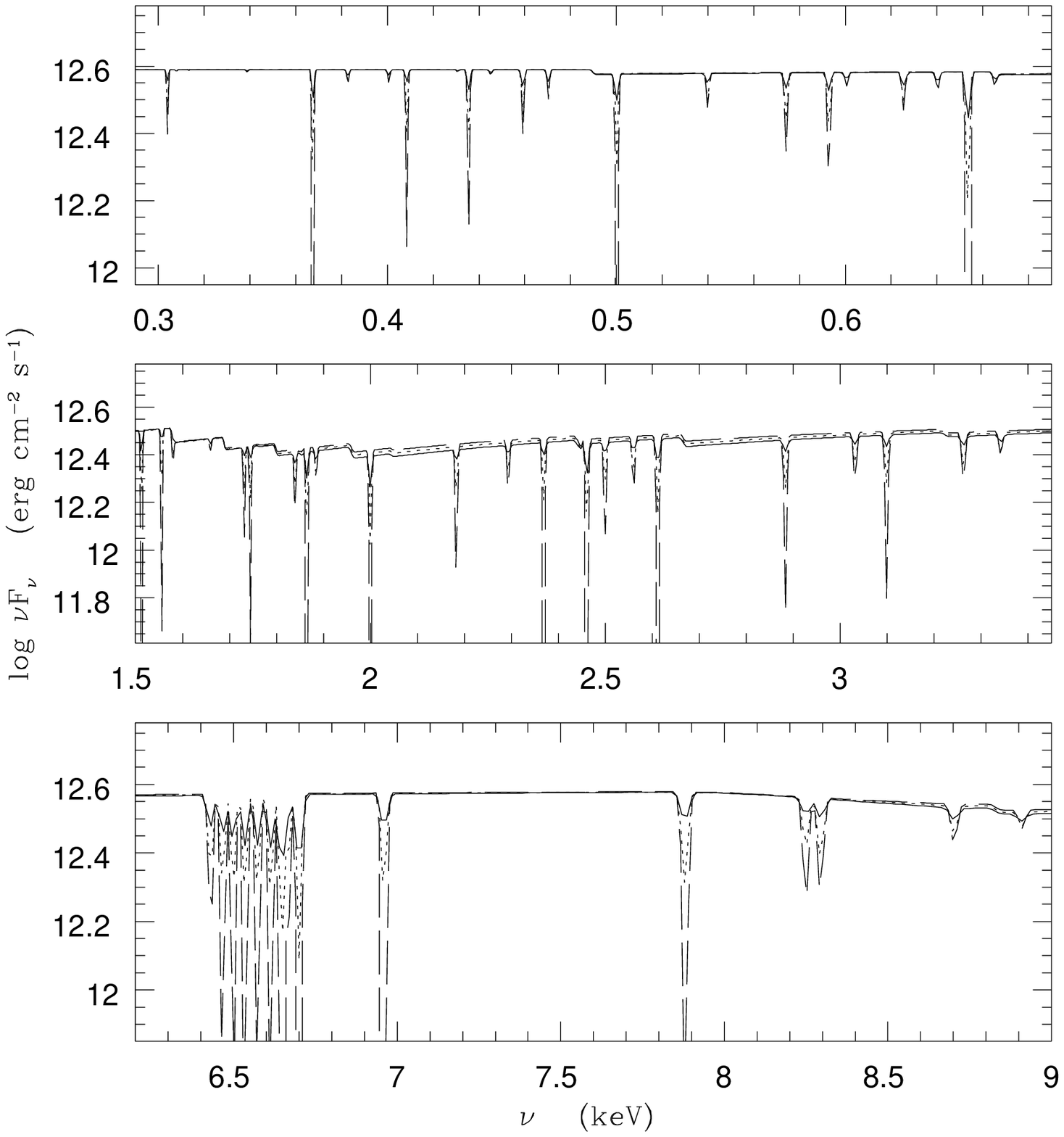}
\caption{Set of transmission spectra in case
of turbulent velocity for  $\Gamma = 2.0$, $\xi = 10^4$ 
and maximum column density $N_{Htot}^{Max}$.  
Solid line represents the case with $v_{turb}=0$, dotted line
with $v_{turb}=100$ km/s, and dashed line with  $v_{turb}=300$ km/s .}
\label{fig:turb2}
\end{figure}

Therefore, for one specific case  ($\Gamma = 2.0, \xi = 10^4$) we calculate
the models with the maximum value of $N_{Htot}^{Max}$ assuming two values
of the turbulent velocity: 100 km/s and 300 km/s. 
The turbulence affects the thermal structure of the cloud 
(see Fig.~\ref{fig:turb1}). The interior of the turbulent cloud is hotter
than the interior of a 'static' cloud, and cloud interior is less dense.
The continuum is slightly modified in the 2 - 3 keV range but absorption 
lines are naturally much more intense (see Fig.~\ref{fig:turb2}).

Quantitative results for the line
intensities are shown in 
Tab.~\ref{tab:linie1}. We see that the line equivalent widths increase
by a factor of a few. The same trend was already seen in
a constant pressure models of
\Agata et al. (2004). Higher value of the turbulent velocity even 
generally leads
to overprediction of the line strength. However, some predicted 
iron lines are still definitively weaker than in the data. 

We plan to consider more systematically the effects of turbulent velocities 
in our future work.
Metal  abundance is another free parameter in our model which 
should be studied more carefully in the next paper.


\section{Influence of the warm absorber on the relativistic iron line}
\label{sec:rel}

\begin{figure}
\epsfxsize=8.8cm \epsfbox[80 220 580 700]{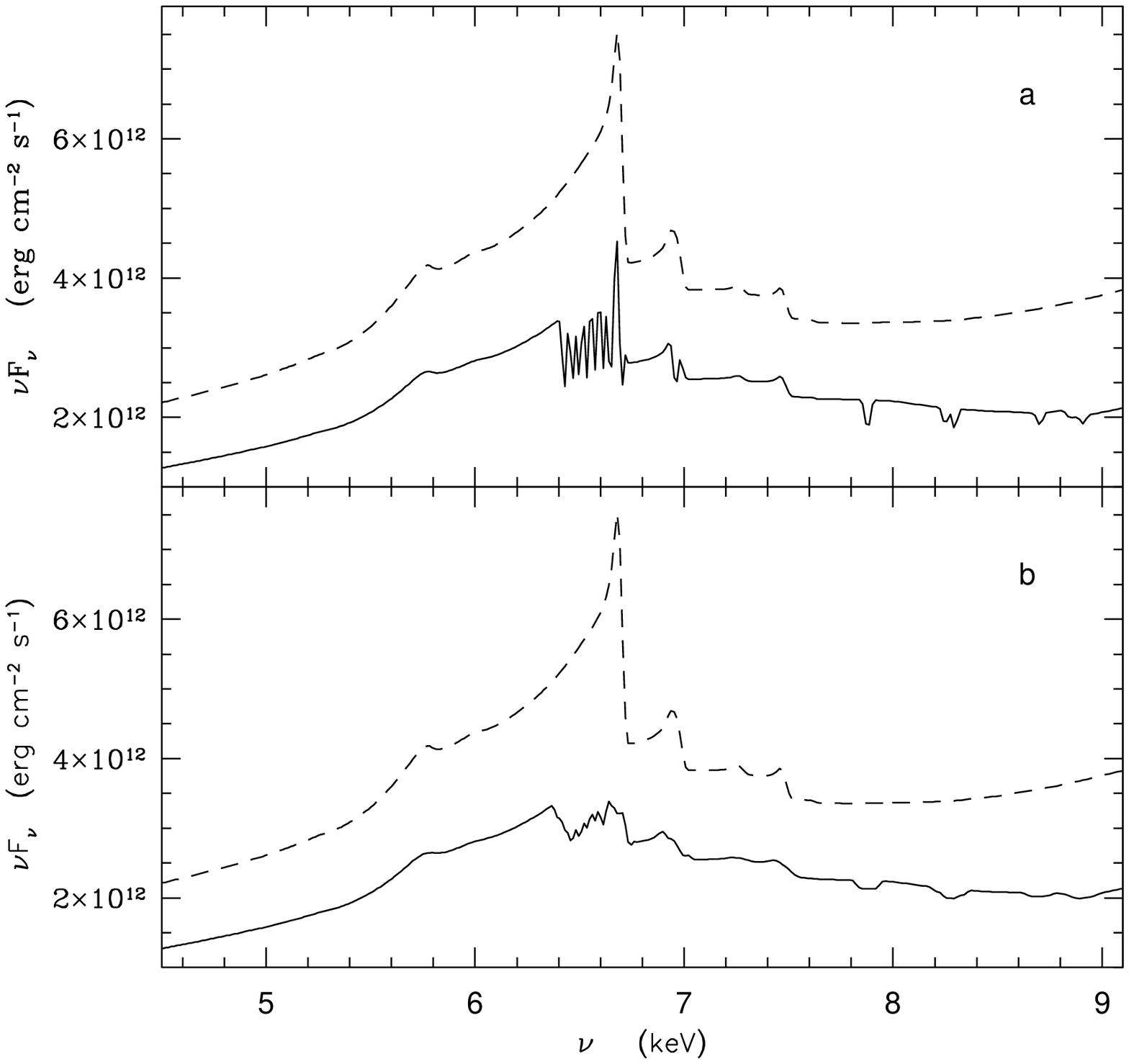}
\caption{An example of the effect of absorption onto the shape of a 
relativistically smeared disk line. The unabsorbed disk spectrum 
is marked by dashed line, while the solid line represents
spectrum with two resolutions: 
a resolution of 390 in panel a, 
and the resolution of {\it Chandra HEG} at 6 keV i.e. 200 in panel b.}
\label{fig:profile}
\end{figure}

A broad, relativistically smeared iron line was discovered in a number of 
sources, MCG -6-30-15 being the first discovered and the best example of this
phenomenon (Tanaka et al. 1995). It is produced by the reflection of the
 hard X-ray continuum on the disk surface very close to the black hole.
If the radiation originating at the disk passes through  
the warm absorber before it reaches the observer the spectral shape
of the line is modified.

To show this effect, we considered an example of the broad disk line. We
have assumed that the disk is illuminated by a random distribution of 
flares following Czerny et al. (2004). Each flare produces a hot spot below,
and each spot emits locally like an illuminated disk atmosphere. 
The final emissivity of the disk
was computed by integrating over all disk radii. 
The black hole was assumed to be rotating, with a Kerr parameter $a = 0.95$ and 
an inclination
to the observer of $i = 30^{\circ}$. All
effects of general relativity were included. 

The spectral shape of the reflected disk
component from this model (without any effect of the warm absorber) is shown in 
Fig.~\ref{fig:profile} as the dashed line in both a and b panels. 
Since the local reflection  comes from the partially ionized disk surface, 
several components of the iron line complex
are still visible in the resulting spectrum, although they are relativistically
smeared. However, if we see the nucleus through the warm absorber
with a  total column density $N_{H_{tot}}=4.5 \times 10^{23}$ and an
ionization parameter $\xi=10^4$
(second spectrum from the bottom in Fig.~\ref{fig:fe10} in the middle panel), 
the overall 
spectrum is suppressed and the narrow absorption features appear 
superimposed onto the broad emission line profile. 

In panel a of Fig.~\ref{fig:profile} we show the result (solid line)
as calculated with the resolution used in our computations $E/\Delta E = 390$  ,
 while in b panel we show the same result degraded down to the resolution of
$E/\Delta E = 200$ characteristic of the {\it Chandra HEG} instrument at 6 keV.

We see that a strong absorption modifies the overall shape of 
the broad iron line. 
The absorbed shape of line seems to be more symmetric and closer to 
a Gaussian shape. 
Also the fitted representative
energy (mean energy of the line) would be lower than derived for 
the intrinsic spectrum.
The situation is therefore similar to the case of UV spectra where
we see narrow absorption features in BLR lines.

\begin{table*}

\caption{Ratios of Equivalent Widths of lines from the same ion in the  
optically thin 
approximation. For comparison the results from observations are 
given if available. We calculate the 
error in the most pessimistic way, when
the numerator has its maximal value EW+$\Delta$EW, and the denominator has 
its minimal value EW-$\Delta$EW.  \label{tab:ew}}
\begin{center}     
\renewcommand{\arraystretch}{1.0}
\begin{tabular}{lclllll}  
\hline

Name of Ratio  & Theory   &    NGC  3783     &    NGC 5548    &  Mrk 509            
& NGC 4051               &  Ton S180  \\ 
        & $f_{ij}/f_{ik}$ &   Kaspi et al.  &  Kaastra et al. &  Yaqoob et al.  
&  Collinge et al.  &   \Agata et al. \\
        &                & 2002             & 2002      & 2003 & 2001& 2004\\
\hline
C{\sc v}$_{(0.3079)/(0.3545)}$    & 4.6   & --      & $3.85 \pm 23.15$   & -- &--  &--  \\
\hline    
C{\sc vi}$_{(Ly \alpha)/(Ly \beta)}$  &  8.49 & --      & $2.33 \pm 2.99$    & -- &--  &--  \\
\hline
C{\sc vi}$_{(Ly \alpha)/(Ly \gamma)}$ & 14.34 & --      & $3.77 \pm 14.94$   & -- &--  &--  \\
\hline
C{\sc vi}$_{(Ly \alpha)/(Ly \delta)}$ & 29.86 & --      & -- & -- &-- & $0.62^{+3.53}_{-0.59}$ \\
\hline 
N{\sc vii}$_{(Ly \alpha)/(Ly \beta)}$ & 5.25 & $1.53\pm 4.39$ & $12.09\pm 1.18$& $0.86\pm 0.98 $ &-- &--  \\
\hline 
N{\sc vii}$_{(Ly \alpha)/(Ly \gamma)}$  & 14.34 & $3.5\pm 8.3 $  & -- & -- &--  &--  \\
\hline 
N{\sc vii}$_{(Ly \alpha)/(Ly \delta)}$  & 29.92 & $1.48\pm 2.27$ & -- & -- &--  &--  \\
\hline
O{\sc vii}$_{(0.574)/(0.6656)}$  & 4.75  & $0.95\pm 1.76$ & $1.32\pm 0.58$ & -- &$1.11\pm 1.45$ &--  \\
\hline 
O{\sc vii}$_{(0.574)/(0.6977)}$  & 1.26  & $0.73\pm 1.01$ & $2.13\pm 2.21$ & -- &$1.43\pm 2.48$ &--  \\
\hline 
O{\sc viii}$_{(Ly \alpha)/(Ly \beta)}$ & 5.26  & $1.04\pm 0.75$ & $1.16\pm 0.55$ & -- &$2.06\pm 1.37$ & -- \\
\hline 
O{\sc viii}$_{(Ly \alpha)/(Ly \gamma)}$ & 14.34 & $0.89\pm 0.55$ & $ 1.54 \pm1.17$& -- &-- & -- \\
\hline
O{\sc viii}$_{(Ly \alpha)/(Ly \delta)}$ & 29.92 & $0.71\pm 0.41$ & $7.36 \pm 1.46$& -- &-- & -- \\ 
\hline
Mg{\sc xi}$_{(1.352)/(1.579)}$  & 4.84  & $1.15\pm 0.19$ & -- & -- &-- & -- \\             
\hline
Mg{\sc xi}$_{(1.352)/(1.660)}$  & 12.9  & $1.46\pm 0.34$ & -- & -- &-- & -- \\             
\hline
Si{\sc xiii}$_{(1.865)/(2.182)}$& 4.97  & $1.73\pm 0.51$ & -- & -- &-- & -- \\             
\hline
Si{\sc xiii}$_{(1.865)/(2.294)}$& 13.09 & $2.19\pm 1.37$ & -- & -- &-- & -- \\             
\hline
Si{\sc xiv}$_{(Ly \alpha)/(Ly \gamma)}$ & 14.34 & $3.37\pm 2.12$ & -- & -- &-- & -- \\             
\hline
S{\sc xv}$_{(2.461)/(2.883)}$   & 4.76  & $2.48\pm 1.98$ & -- & -- &-- & -- \\             
\hline
S{\sc xv}$_{(2.461)/(3.032)}$   & 13.18 & $1.58\pm 0.95$ & -- & -- &-- & -- \\             
\hline
Fe{\sc xvii}$_{(0.8211)/(0.8984)}$ & 9.51 & $0.47\pm 0.24$ & $0.88\pm 0.99$ & -- &-- & -- \\             
\hline
Fe{\sc xvii}$_{(0.8211)/(1.008)}$ & 3.01 & $0.67\pm 0.22$ & -- & -- &-- & -- \\   
\hline
Fe{\sc xxv}$_{(6.700)/(7.881)}$  & 4.95 & $0.42\pm 0.93$ & -- & -- &-- & -- \\            
\hline     
\end{tabular}
\end{center} 

\end{table*}

\section{Comparison with observations}
\label{sec:comp}

Recent observations of Sy1 or Sy1.5 galaxies show the presence
of absorption lines in soft X-ray spectra from {\it Chandra}
or {\it XMM} satellites. 
The number of detected lines depends on the resolution and
sensitivity of the detector, and the best observations
were obtained for NGC 3783 (Kaspi el al. 2002, Behar el al. 2003,
Netzer et al. 2003, Krongold el al. 2003). A 900 ksec observation 
allowed for the detection of over few tens of lines. 
In the cases of about 90 ksec observations or shorter,  
only a few lines were detected and fitted like in case of Mrk 509,
TonS180, PG1211+143, or NGC 7469 (Blustin et al. 2003).
Therefore,  it is hard  to fit our models 
with those observations, but we 
make some comparison by examining parameters  already fitted by other 
authors.

First of all, the data suggest that the total column density of 
the warm absorber is never higher than  $ \sim  10^{23} $ cm$^{-3}$
(Piconcelli et al. 2004).
In our models we have a physical limit of the column densities 
because of the presence of thermal instabilities. 
For each spectral index and ionization parameter there is a maximum value of
the column density for which warm absorber exists (see Fig.~\ref{fig:grid}). 
In all cases the values of this maximum total column densities
agree with the column densities determined from observations 
of different object. 

As we mentioned above, in a stratified medium it is 
difficult to determine the plasma parameters by comparing directly
the EWs of 
individual lines. Therefore, in Tab.~\ref{tab:ew}
we present the ratios of the EWs of lines for the same ion.
In the first column names of ratios are given, 
in the second column those ratios are computed theoretically 
assuming that lines are optically thin (see formula 2. 
in  \Agata et al. 2004). We have listed only those line ratios
which were detected by X-ray satellites. Columns from 3 up to 7
present the observed ratios assuming the most pessimistic error, when
the numerator has its maximal value, and the denominator has 
its minimal value (see Tab.~\ref{tab:ew}).

The observed line ratios  are usually much lower than predicted
in the optically thin medium limit.  
Therefore, observations strongly suggest that
lines are optically thick and radiative transfer is required to 
predict equivalent widths. 

The same ratios computed for our models are presented
in Tab.~\ref{tab:ewmod}. For each set of models with spectral index
and ionization parameter, the ratios are of the order 
of those observed  when we consider clouds with  higher total
column densities.  
We conclude here that when the warm absorber have a  
total column density higher than $10^{22}$ cm$^{-2}$ the resonance lines
of the main ions
become optically thick and radiative transfer 
should be used to model them.  
For instance, the most popular $Ly_{\alpha}$ lines of: 
O{\sc viii}, Ne{\sc x}, Mg{\sc xii}, Si{\sc xiv}, and S{\sc xvi} 
for $\xi=10^4$ $\Gamma=2.0$ and log$N_{H_{tot}}$=23.5 
have optical depths integrated on the whole cloud equal respectively:
$977$,  $ 363 $, $234 $,
$ 436 $, and  $216 $.






\section{Discussion}
\label{sec:disc}

We have calculated a grid of models of warm absorber with density stratification
determined by the condition of the constant
pressure.
The assumption of constant pressure instead of constant density was
motivated by recent observations of AGN. 
In most observed objects the ionization states implied by the observed lines 
span a large range, which cannot be accounted for a
single photoionized region, or by collisionally ionized matter. At least
two photoionized absorbing regions are required to fit the data
and the properties of these two regions are consistent with pressure equilibrium 

Our models represent an essential improvement in the 
description of a single cloud over the models presented in the 
literature. All previous models adopted constant density approximation 
for the medium (Netzer 1993, Kaastra et al. 2002, Kinkhabwala et al. 2002,
Krongold et al. 2005), or even constant temperature  (Krolik \&
Kriss 1995).  On the other hand, we do not consider here a global picture
including the dynamics of the warm absorber flow which form a separate and
a major issue (see Chelouche \& Netzer 2005 and the references therein).
 
We have calculated the full radiative transfer of the illuminating radiation
in continuum and in lines through the plane-parallel density-stratified 
slabs of different total column densities. 
For a low total column density the cloud is mostly ionized and optically thin, 
and does not differ from the constant density model. But for a higher
total column density the optically thick dense zone arises at the
back of the illuminated cloud, and the temperature falls dramatically. 
Therefore, absorbing matter contains zones of different
ionization states coexisting under constant pressure.

For each set of models with spectral index $\Gamma$ and surface ionization
parameter $\xi$ there is a maximum total column density
$N^{Max}_{H_{tot}}$ for which the cloud is thermally stable. 
For higher column densities thermal instabilities 
do not allow the calculations to converge. 
Interestingly, $N^{Max}_{H_{tot}}$  is of the order of 
the maximum column density derived from X-ray observations 
of different AGN. It gives additional support to the idea
of the warm absorber clouds being in pressure equilibrium.  

The modeled lines are usually optically thick and their
equivalent widths are of the order of
the observed values. We conclude that the observed
warm absorbers at column densities of the order of
$10^{22}$ cm$^{-2}$ and higher possess saturated absorption 
lines and that full radiative transfer is required to 
model their equivalent widths properly. 
For full evaluation  of this problem we will perform a curve of growth 
analysis in our future work. 

Generally lines are easier to detect than ionization edges and
this tendency is observed in several AGN (Kaastra et al. 2002,
 \Agata et al. 2004). 
Our models predict that edges are not observable up to 
column densities about $10^{22}$ cm$^{-2}$ and even higher, depending
on $\Gamma$ and $\xi$.

The most interesting result of our computations is the shape 
of the spectrum around an iron K$\alpha$ line. 
In almost all models there are strong and narrow absorption 
features due to highly ionized iron ions, above Fe{\sc xvii}. 
Such absorption affects the shape of the broad iron emission line 
possibly originating in illuminated disk atmosphere. 
For instance, relativistic broad iron line profile after passing 
through warm absorber becomes disrupted into three narrower
profiles, which can be fitted by Gaussian shape. 
Such lines were reported in several AGN as
presented by Yaqoob \& Padmanabhan (2004) (see also NGC 3783  Reeves et al. 2004). 

In this paper we consider only single cloud but observations
suggest that absorbing material forms distribution of clouds with 
covering factor which may depend on velocity of clouds (de Kool, 
Korista \& Arav 2002). However, theoretical modeling 
of cloud distribution with complex velocity field is 
complicated.
Also, the study of the transmission spectra, including absorption lines, do not 
give us any diagnostic of the 
matter along other directions than the line of sight. The {\sc titan} 
code can be 
used, however, to study the radiative transfer in all directions. 
We address both issues to the future work.

\begin{acknowledgements}

We thank to Suzy Collin, Martine Mouchet and
Aneta Siemiginowska for helpful discussion. 
Part of this work was supported by grants 2P03D00322 and 
1P03D00829 of the Polish
State Committee for Scientific Research,
the Laboratoire Europe\' en Associ\' e Astrophysique Pologne-France,
and by the Hans-B{\"o}ckler-Stiftung.

\end{acknowledgements}

\begin{landscape}
\begin{table}

\caption{Ratios of Equivalent widths of lines from the same ion
computed for our models. We have chosen two column densities
for each set of $\Gamma$ and $\xi$ where  more less transition from
optically thin to optically thick cloud occurs  
\label{tab:ewmod}.}
\renewcommand{\arraystretch}{1.1}
\begin{tabular}{|c|r|r|r|r|r|r|r|r|r|r|r|r|r|r|r|r|r|r|}  
\hline
$\Gamma$  & \multicolumn{6}{|c|}{1.5} & \multicolumn{6}{|c|} {2.0}  & \multicolumn{6}{|c|}{2.5}    \\
\hline 
$\xi$  &\multicolumn{2}{|c|}{$10^5$} & \multicolumn{2}{|c|} {$10^4$} & \multicolumn{2}{|c|}{$10^3$} & 
        \multicolumn{2}{|c|}{$10^5$} & \multicolumn{2}{|c|} {$10^4$} & \multicolumn{2}{|c|}{$10^3$} & 
        \multicolumn{2}{|c|}{$10^5$} & \multicolumn{2}{|c|} {$10^4$} & \multicolumn{2}{|c|}{$10^3$}  \\
\hline 
$log(N_{H_{tot}})$        & 23.7 & 23.8 & 21.  & 22.5  &  22. & 22.6 & 21.5 & 23.5 & 21.5 & 23.5 & 21.5 & 23.5 & 21.5 & 23.5 & 21.5 & 23.4 & 21.5 & 23.2 \\
\hline
\hline
C{\sc v}$_{(0.3079)/(0.3545)}$    & 4.57 & 1.12 & 1.27 & 3.88  & 4.49 & 3.32 & 4.61 & 4.61 & 4.61 & 4.56 & 4.61 & 1.34 & 4.60 & 1.72 & 4.61 & 2.15 & 4.61 & 2.40 \\
\hline 
C{\sc vi}$_{(Ly \alpha)/(Ly \beta)}$   & 5.09 & 2.04 & 1.24 & 1.55  & 1.32 & 1.40 & 5.27 & 2.54 & 5.25 & 1.15 & 3.52 & 1.16 & 5.19 & 1.19 & 4.68 & 1.16 & 2.13 & 2.03 \\
\hline 
C{\sc vi}$_{(Ly \alpha)/(Ly \gamma)}$   & 13.64& 2.23 & 1.20 & 2.03  & 2.17 & 1.67 & 14.35& 6.06 & 14.3 & 1.35 & 9.05 & 1.31 & 14.37& 1.39 & 12.77& 1.56 & 4.97 & 2.45 \\
\hline 
C{\sc vi}$_{(Ly \alpha)/(Ly \delta)}$   & 28.61& 2.30 & 1.17 & 2.62  & 3.81 & 2.11 & 29.89& 12.12& 29.79& 1.69 & 18.54& 1.52 & 29.65& 1.61 & 26.26& 1.57 & 9.75 & 2.80 \\
\hline 
N{\sc vii}$_{(Ly \alpha)/(Ly \beta)}$  & 5.25 & 1.73 & 0.97 & 1.35  & 1.64 & 1.26 & 5.26 & 3.12 & 5.27 & 1.18 & 4.08 & 117 & 5.23 & 1.24 & 4.87 & 1.20 & 2.58 & 1.86\\
\hline 
N{\sc vii}$_{(Ly \alpha)/(Ly \gamma)}$  & 13.91& 1.75 & 1.09 & 1.82  & 3.23 & 1.59 & 14.36& 7.8  & 14.34& 1.46 & 10.76& 1.36 & 14.64& 1.50 & 13.50& 1.45 & 6.40 & 2.27\\
\hline 
N{\sc vii}$_{(Ly \alpha)/(Ly \delta)}$  & 29.61& 1.82 & 1.61 & 2.32  & 6.25 & 2.18 & 29.89& 15.88& 28.57& 1.99 & 22.18& 1.62 & 29.79& 1.74 & 27.38& 1.69 & 12.57& 2.56\\
\hline
O{\sc vii}$_{(0.574)/(0.6656)}$  & 4.80 & 3.55 & 2.96 & 2.95  & 2.75 & 3.02 & 4.76 & 4.63 & 4.77 & 1.76 & 4.68 & 2.17 & 4.74 & 2.26 & 4.75 & 1.88 & 4.56 & 2.08\\
\hline 
O{\sc vii}$_{(0.574)/(0.6977)}$  & 12.83& 4.96 & 3.39 & 5.06  & 6.66 & 6.06 & 12.66& 12.26& 12.67& 3.62 & 12.39& 2.93 & 12.54& 2.90 & 12.56& 2.31 & 9.41 & 5.87\\
\hline 
O{\sc viii}$_{(Ly \alpha)/(Ly \beta)}$ & 4.61 & 2.57 & 1.37 & 1.29  & 1.14 & 1.38 & 5.27 & 1.43 & 5.14 & 1.47 & 1.47 & 1.82 & 5.14 & 2.19 & 2.40 & 2.01 & 1.18 & 2.57\\
\hline 
O{\sc viii}$_{(Ly \alpha)/(Ly \gamma)}$ & 12.47& 3.88 & 133 & 1.50  & 1.34 & 1.52 & 14.35& 2.14 & 13.96& 1.64 & 2.74 & 1.95 & 13.86& 2.38 & 5.64 & 2.16 & 1.50 & 2.92\\
\hline
O{\sc viii}$_{(Ly \alpha)/(Ly \delta)}$ & 25.52& 4.60 & 1.36 & 1.67  & 1.62 & 1.64 & 29.89& 3.44 & 29.02& 1.79 & 5.03 & 2.07 & 29.16& 2.59 & 11.38& 2.35 & 2.18 & 3.18\\
\hline
Mg{\sc xi}$_{(1.352)/(1.579)}$  & 4.85 & 1.89 & 2.82 & 1.64  & 2.42 & 1.32 & 487 & 4.34 & 5.36 & 1.43 & 4.49 & 1.72 & 4.94 & 2.27 & 4.88 & 2.15 & 2.17 & 2.59\\            
\hline
Mg{\sc xi}$_{(1.352)/(1.660)}$  & 13.05& 1.95 & 6.90 & 2.90  & 5.66 & 2.28 & 13.0 & 11.36& 14.33& 1.98 & 11.84& 2.00 & 13.11& 2.61 & 12.89& 2.46 & 4.75 & 3.06\\          
\hline
Si{\sc xiii}$_{(1.865)/(2.182)}$& 4.99 & 1.62 & 4.72 & 1.55  & 1.63 & 1.62 & 5.02 & 2.82 & 5.05 & 2.04 & 3.26 & 2.34 & 5.14 & 2.99 & 4.28 & 2.85 & 1.64 & 2.64\\            
\hline
Si{\sc xiii}$_{(1.865)/(2.294)}$& 13.05& 2.28 & 12.2 & 2.25  & 2.88 & 1.98 & 13.23& 6.64 & 13.32& 2.34 & 7.96 & 2.63 & 13.64& 3.47 & 11.03& 3.29 & 2.75 & 3.05\\            
\hline
Si{\sc xiv}$_{(Ly \alpha)/(Ly \gamma)}$ & 13.10& 7.18 & 14.43& 2.06  & 2.24 & 1.66 & 14.41& 2.64 & 14.09& 2.11 & 5.66 & 2.11 & 13.84& 2.26 & 5.89 & 2.18 & 3.52 & 1.79\\        
\hline
S{\sc xv}$_{(2.461)/(2.883)}$   & 4.86 & 2.20 & 5.05 & 1.66  & 1.76 & 1.72 & 4.82 & 2.36 & 4.89 & 2.39 & 2.96 & 2.47 & 4.89 & 2.90 & 3.59 & 2.77 & 1.83 & 2.36\\           
\hline
S{\sc xv}$_{(2.461)/(3.032)}$   & 13.14& 4.83 & 13.7 & 2.64  & 3.27 & 2.21 & 13.36& 5.47 & 13.51& 2.79 & 7.42 & 2.84 & 13.93& 3.48 & 9.66 & 3.32 & 3.68 & 2.81\\           
\hline
Fe{\sc xvii}$_{(0.8211)/(0.8984)}$& --   & 4.15 & 10.8 & 3.9   & 5.21 & 3.68 & --   & 11.5 & --   & 4.12 & 11.87& 4.12 & --   & 4.77 & 12.29& 4.32 & 5.59 & 3.03\\           
\hline
Fe{\sc xvii}$_{(0.8211)/(1.008)}$& --   & 2.07 & 3.20 & 1.95  & 1.93 & 1.90 & --   & 3.39 & --   & 1.83 & 3.43 & 1.76 & --   & 1.65 & 3.51 & 1.54 & 2.00 & 1.18\\ 
\hline
Fe{\sc xxv}$_{(6.700)/(7.881)}$ & 4.90 & 3.82 & 5.20 & 4.11  & 3.95 & 3.4  & 5.12 & 2.00 & 4.71 & 2.52 & 3.76 & 2.4  & 3.86 & 2.60 & 2.77 & 2.49 & 5.92 & 5.33\\
\hline
                
\end{tabular}
\end{table}

\end{landscape}


\begin{thebibliography}{99}

\bibitem[]{} Ashton A.C., Page M.J., Branduardi-Raymont G., 
         Blustin A.J., 2005, astro-ph/0511748
\bibitem[]{} Antonucci, R. R. J., \& Miller, J. S. 1985, ApJ, 297, 621
\bibitem[]{} Ballantyne, D. et al., 2003, A\&A, 409, 503
\bibitem[]{} Barcons, X., Carrera, F.J., Ceballos, M.T., 2003, MNRAS, 346, 897
\bibitem[]{} assani, L., Dadina, M., Maiolino, R., Salvati, M., Risaliti, G., 
 della Ceca, R., Matt, G., Zamorani, G., 1999, ApJS, 121, 473
\bibitem[Behar et al.(2003)]{2003ApJ...598..232B} Behar, E., Rasmussen, 
A.~P., Blustin, A.~J., Sako, M., Kahn, S.~M., Kaastra, J.~S., 
Branduardi-Raymont, G., \& Steenbrugge, K.~C.\ 2003, \apj, 598, 232 
\bibitem[]{} Blustin, A.J. et al. 2003, A\&A, 403, 481
\bibitem[]{} Blustin et al. 2005, A\&A, 431, 569
\bibitem[]{} Chartas, G., Brandt, W.N., Gallagher, S.C., Garmire, G.P., 2002, 
      ApJ, 279, 169  
\bibitem[]{} Chelouche D., Netzer H., 2005, ApJ, 625, 95
\bibitem[]{} Collin, S., Dumont, A.-M., Godet, O., 2004, A\&A, 
\bibitem[]{} Collinge, M.J. et al.,  2001, ApJ, 557, 2
\bibitem[]{} Crenshaw, D.M., Kraemer, S.B. \& George, I.M., 2003a,
  Annu. Rev. Astron. Astrophys., 41,117 
\bibitem[]{} Crenshaw, D.M. et al., 2003b, ApJ, 594, 116
\bibitem[]{} Czerny, B., \Agata, A., Dovciak, M., Karas, V., Dumont, A.-M., 
      2004, A\&A, 419, 877
\bibitem[]{} Dadina, M., Cappi, M., 2004, A\&A, 413, 921 
\bibitem[]{} Dasgupta S., Rao A.R., Dewangan G.C., Agrawal V.K., 2005, 
             ApJ, 618, L87
\bibitem[]{} de Kool M., Korista K.T., Arav, N. 2002, ApJ, 580, 54
\bibitem[]{} Dumont, A.-M., Abrassart A., Collin S., 2000, A\&A, 357, 823
\bibitem[]{} Dumont, A.-M., Collin S., Paletou F., Coupé, S., Godet O., 
      Pelat, D. 2003, A\&A, 400, 437
\bibitem[]{} Field G.B., 1965, ApJ, 142, 531
\bibitem[]{} Gabel J.R. et al., 2005, ApJ, 631, 741
\bibitem[]{} George I.M., Turner, T. J., Netzer, H., Nandra, K., 
      Mushotzky, R. F., Yaqoob, T. 1998, ApJS, 114, 73
\bibitem[]{} Guainazzi, M., Fiore, F., Matt, G., Perola, G. C., 2001, 
            MNRAS, 327, 323
\bibitem[]{} Halpern J.P., 1984, ApJ, 281, 90
\bibitem[]{} Kaastra, J.S., Steenbrugge, K.C., Raassen, A.J.J., 
      van der Meer, R.L.J., Brinkman, A.C., Liedahl, D.A., Behar, E., 
      de Rosa, A.,
      2002, A\&A, 386, 427
\bibitem[]{} Kaspi S., 2004a, in "The Interplay among Black Holes, Stars and 
     ISM in Galactic Nuclei", IAU Symposium 222, eds. Th. Storchi Bergmann, 
     L.C. Ho \& H.R.; p.41-44
\bibitem[]{} Kaspi, S., Brandt, W. N., Netzer, H., George, I.M., Chartas, G., 
      Behar, E., Sambruna, R.M., Garmire, G.P., Nousek, J.A., 2001, ApJ, 
      554, 216
\bibitem[Kaspi et al.(2002)]{2002ApJ...574..643K} Kaspi, S., et al.\ 2002, 
\apj, 574, 643 
\bibitem[]{} Kaspi, S., Netzer, H., Chelouche, D., George, I.M., 
         Nandra, K., Turner, T.J., 2004b, ApJ, 611, 68
\bibitem[]{} Kinkhabwala, A. et al. 2002, ApJ, 575, 732

\bibitem[]{} Krolik, J. 2002, ``Workshop on X-ray Spectroscopy of AGN with Chandra 
     and XMM-Newton'', MPE Report 279, p. 131
\bibitem[]{} Krolik, J., Kriss, G.A., 1995, ApJ, 447, 512
\bibitem[]{} Krolik, J., Kriss, G.A., 2001, ApJ, 561, 684
\bibitem[Krolik, McKee, \& Tarter(1981)]{1981ApJ...249..422K} Krolik, 
J.~H., McKee, C.~F., \& Tarter, C.~B.\ 1981, \apj, 249, 422 
\bibitem[]{} Krongold, Y., Nicastro, F., Brickhouse, N.S., Elvis, M., 
     Liedahl, D.A., Mathur, S., 2003, ApJ, 597, 832
\bibitem[]{} Krongold, Y., Nicastro, F., Brickhouse, N.S., Elvis, M., 
        Mathur, S., 2005, ApJ, 622, 842
\bibitem[]{} Matt G. , 2000, A\&A, 327, L31
\bibitem[Matt, Bianchi, D'Ammando, \& 
     Martocchia(2004)]{2004A&A...421..473M} Matt, G., Bianchi, S., D'Ammando, 
      F., \& Martocchia, A.\ 2004, \aap, 421, 473 
\bibitem[]{} Mushotzky, R.F., Done, C., Pounds, K.A., 1993, ARA\&A, 31, 717
\bibitem[Nandra \& Pounds(1992)]{1992Natur.359..215N} Nandra, K.~\& Pounds, 
           K.~A.\ 1992, \nat, 359, 215 
\bibitem[]{} Nandra K., \& Pounds, K.A., 1994, MNRAS, 268, 405
\bibitem[]{} Netzer, H., 1993, ApJ, 411, 594

\bibitem[]{} Netzer, H., 1996, ApJ, 473, 781

\bibitem[]{} Netzer, H., Chelouche, D., George, I.M., Turner, T. J., 
   Crenshaw, D. M., Kraemer, S. B., Nandra, K., 2002, ApJ, 571, 256
\bibitem[Netzer et al.(2003)]{2003ApJ...599..933N} Netzer, H., et al.\ 
2003, \apj, 599, 933 

\bibitem[]{} Otani, C. et al., 1996, PASJ, 48, 211
\bibitem[]{} Piconcelli, E., Jimenez-Bail{\'o}n, E., Guainazzi M., Schartel N., 
  Rodr{\'i}guez-Pascual P.M. and Santos-Lle{\'o} M., 2004, MNRAS, 351, 161
\bibitem[]{} Pounds, K.A., King, A.R, Page, K.L., O'Brien, P.T., 2003a, MNRAS, 
       346, 102 
\bibitem[]{} Pounds K.A.,  Page K.L., 2005, MNRAS, 360, 1123
\bibitem[]{} Pounds, K.A.,  Reeves, J. N., King, A. R., Page, K. L., 
    O'Brien, P. T., 
    Turner, M. J. L., 2003b, MNRAS, 346, 1025
\bibitem[]{} Reeves, J.N., Nandra, K., George, I.M., et al., 2004, ApJ, 
     602, 648
\bibitem[]{} Reynolds C.S., 1997, MNRAS, 286, 513
\bibitem[]{} Reynolds C.S., Fabian A.C., Nandra K., Inoue H., Kunieda H., 
     Iwasawa K., 1995, MNRAS, 277, 901
\bibitem[]{} \Agata A., Czerny B., 1996, Ac. A., 46, 233
\bibitem[]{} \Agata A., Czerny B., Dumont A.-M., Collin S., Siemiginowska A.,
     2004, ApJ, 600, 96
\bibitem[]{} R{\' o}{\. z}a{\' n}ska, Dumont, Czerny, \& 
        Collin(2002)]{2002MNRAS.332..799R} R{\' o}{\. z}a{\' n}ska, A., Dumont, 
        A.-M., Czerny, B., \& Collin, S.\ 2002, \mnras, 332, 799 
\bibitem[]{} Scott J.E., Kriss, G.A., Lee, J.C., Quijano, J.K., 
      Brotherton, M. et al., 2005, ApJ, 634, 193
\bibitem[]{} Steenbrugge K.C., Kaastra J.S., Sako M., Branduardi-Raymont G., 
             Behar E. et al., 2005a, A\&A, 432, 453
\bibitem[]{} Steenbrugge K.C., Kaastra J.S., Crenshaw D.M., Kraemer S.B. 
             Arav N. et al., 2005b, A\&A, 434, 569
\bibitem[]{} Schurch N.J., Warwick R.S., Griffiths R.E., Kahn S.M., 
 2004, MNRAS, 350, 1
\bibitem[]{} Shih D.C., Iwasawa K., Fabian A.C., 2003, MNRAS, 341, 973
\bibitem[]{} Tanaka Y., et al. 1995, Nature, 375, 659
\bibitem[]{} Turner A.K., Fabian A.C., Lee J.C., Vaughan S., 2004, MNRAS, 
     353, 319
\bibitem[]{} Turner T.J, George I.M., Nandra K., Mushotzky R.F., 1997, ApJS, 113, 23
\bibitem[]{} Weaver, K.A., Reynolds, C.S., 1998, ApJ, 503, L39
\bibitem[]{} Worsley et al. 2004, MNRAS, 350, 207
\bibitem[]{} Yaqoob, T., McKernan, B., Kraemer, S. B., Crenshaw, D. M.,
     Gabel, J. R., George, I. M.,Turner, T. J. 2003, ApJ, 582, 105
\bibitem[]{} Yaqoob, T., Padmanabhan, U., 2004, ApJ, 604, 63 

\end{thebibliography}
\end{document}